# Thermal spin transport and spin in thermoelectrics

Joseph P. Heremans (first name to appear in full)

Department of Mechanical and Aerospace Engineering, Department of Material Science and Engineering and Department of Physics, The Ohio State University, Columbus, OH 43210, USA

**Abstract**

This article reviews the principles that govern the combined transport of spin, heat, and charge, both from a macroscopic point of view (the Onsager relations) and microscopically (transport by spin-polarized electrons and magnons). The extensive thermodynamic quantity associated with spin transport is the magnetization; its Onsager-conjugate force is in general the derivative of the free energy with respect to the magnetization. The spin-angular momentum is uniquely associated with the magnetization, so that the words "spin" and "magnetization" are used interchangeably. Spins are carried in one of two ways: (1) by spin-polarized free electrons in magnetic metals and doped semiconductors, or (2) by spin waves (magnons) that reside on localized electrons on unfilled *d*- or *f*-shells of transition metal or rare-earth elements. The paper covers both cases in separate chapters. In both cases, it is possible to define a spin chemical potential whose gradient is the more practical conjugate force to spin transport. The paper further describes the anomalous Hall, spin Hall, and inverse spin Hall effects in magnetic and non-magnetic solids with strong spin-orbit coupling because these effects are used to generate and measure spin fluxes. Spin transport across interfaces is described next, and includes spin pumping and spin transfer torque. The final chapter then puts all these concepts together to describe the spin-Seebeck, spin-Peltier, and magnon-drag effects, which exist in ferromagnetic, antiferromagnetic, and even paramagnetic solids. Magnon-drag, in particular, is a high-temperature effect that boosts the thermopower of metals by an order of magnitude and that of semiconductors by a factor of 2 or 3 above the electronic diffusion thermopower. This is the only example where a spin-driven effect is larger than a charge-driven effect. Magnon drag leads a simple binary paramagnetic semiconductor, MnTe, to have $zT \geq 1$ without any optimization. This shows how adding spin as an additional design parameter in thermoelectrics research is a new and promising approach toward the quest for high-$zT$ materials.

## 1. Introduction

Thermal spin transport concerns the mixed transport of heat and spin, or, more precisely, magnetic moment, just as thermoelectric transport is concerned with the mixed transport properties of heat and electrical charge. Thermoelectric research has struggled for long to overcome the counter-indicated nature of the classical transport properties, namely the Seebeck coefficient, the thermoelectric power, and the thermal conductivity, that constitute the *zT*, the thermoelectric figure of merit that covers thermoelectric conversion efficiency. Adding spin to the number of controllable variables adds a new design parameter that inevitably must lead to a better optimum *zT*. Explaining how is the purpose of this paper. The field of thermal spin transport, or *spin caloritronics*, is actually quite old, manifesting mainly by magnon drag identified in ferromagnetic (FM) transition metals like Fe[1] and antiferromagnetic (AFM) semiconductors like MnTe[2] half a century ago. However, the discovery of the spin-Seebeck effect (SSE) on Permalloy in 2008[3] has started a resurgence of the field. Here, we attempt to give a self-contained didactic review, and refer the reader to the numerous review articles[4,5] enumerating more exhaustively the effects involved and the details of the theories in spin-caloritronics.

| Extensive | Name | Charge | Spin, moment | Heat | Number |
|---|---|---|---|---|---|
| | Symbol | $C$ | $\hbar\vec{S}$ , $\vec{\mu}_m = -g\mu_B\vec{S}$ | $Q$ | # |
| | Units | Coulomb | $\hbar$ , $\mu_B$ | Joules | |
| Flow | Name | Current | Spin current | Heat current | particle current |
| | Symbol | $I$ | $I_S$ | $I_Q$ | $I_\#$ |
| | Units | Ampere | $\hbar / s$ , $\mu_B/s$ | Watt | 1/s |
| Flux | Name | Current density | Spin flux | Heat flux | Particle flux |
| | Symbol | $\vec{j}_C$ | $\vec{j}_S$ | $\vec{j}_Q$ | $\vec{j}_\#$ |
| | Units | A $m^{-2}$ | $\hbar s^{-1}m^{-2}$ ,$\mu_B s^{-1}$ $m^{-2}$ | W $m^{-2}$ | $m^{-2}$ |
| Potential | Name | Electrochemical potential | Spin chemical potential | Temperature | Chemical potential |
| | Symbol | $\mu$ | $\mu_S$ | $T$ | $\mu$ |
| | Units | eV | eV | K | eV |
| Conjugate force | Name | Electric field | Spin potential gradient | Temperature gradient | |
| | Symbol | $\vec{E} = -\nabla\mu / e$ | $\nabla\mu_S$ | $\nabla T$ | $\nabla\mu$ |
| | Units | V/m | eV/m | K | eV/m |

Table 1 Thermodynamic quantities for combined charge, spin and heat transport.

The flow of any well-defined thermodynamic quantity based on a physical, observable effect, along with its conjugate force, obeys Onsager reciprocity. The Onsager relations describe the effect on a flux of an extensive thermodynamic quantity, here charge *C*, heat *Q*, and magnetic moment *M* or spin, of thermodynamic forces, which themselves are gradients of potentials (intensive thermodynamic variables). Table 1 gives an overview of the quantities involved. The flux of charge is the current density $\vec{j}_C$, and so on with spin and heat. The direct thermodynamic force that generates charge flow is $\vec{F} = e\vec{E}$, where *e* is the charge of the electron ($e$=1.6×10$^{-19}$ C), the electric field $\vec{E} = -\nabla\mu / e$ being itself the gradient of the electrochemical potential $\mu$. In heat transport, the heat is the extensive quantity, and its flux $\vec{j}_Q$ is driven by its conjugate force, the temperature gradient $\nabla T$. Relations between fluxes and thermodynamic forces are the Onsager relations, and, in most cases, are assumed to be linear.

Spin transport formally is treated the same way as charge and heat transport, and the Onsager relations will be extended here to include it. The most important thermodynamic quantity is the magnetization itself, the quantity whose transport is considered in this paper. The notation used for magnetization or magnetic moment (magnetization per unit volume) is as follows: $\vec{M}$ is the total magnetization of the sample, $\vec{m}$ is the moment per unit volume, and $\vec{\mu}$ is the moment per atom. The most convenient unit used to express the moment is the Bohr magneton $\mu_B = e\hbar/2m = 5.788\ 10^{-5}$ eV/Tesla. The spin-angular momentum on each atom is $\hbar\vec{S}$. The magnetic moment of each atom is then:

$$\vec{\mu} = -g\mu_B\vec{S} \qquad (1.1),$$

where g is the Landé factor, typically 2. The same equation also relates the time derivatives of moment and spin-angular momentum, and, thus, also the spin flux $\vec{j}_S$ and the flux of magnetization. Therefore, we use the words spin flux and magnetization flux interchangeably.

There are three distinct ways to carry a flux of magnetization or spin across a sample:

(1) In metals and semiconductors, the free electrons that carry charge and heat in the sample come in either spin-up or spin-down flavor. In non-magnetic material, which we label a *normal metal* (NM), the densities of both are equal. In spin-polarized materials, e.g., FM metals, there are more electrons with their spins oriented parallel to the net magnetization. When this is the case, charge transport is accompanied by spin transport. The thermoelectric effect in mixed charge and heat transport are also accompanied by what is known as *spin-dependent Seebeck and Peltier effects* This will be treated in section 2.

(2) Spin waves exist in FM solids, both FM metals and FM insulators, and in AFMs. They are precessions of the magnetization that resides on the unfilled *d* and *f* levels of the core electrons. Magnon propagation carries both heat and spin fluxes, but no charge flux. This will be treated in section 3. However, magnons can interact with free electrons and transfer their momentum to them, giving rise to an advective transport process called *magnon drag* (MD), that greatly boosts the thermopower of the materials affected and increases their thermoelectric figure of merit *zT*.[6,7,8] Furthermore, magnons can spin-polarize conduction electrons in a NM across an interface between an FM and the NM, by a process called *spin pumping* described in section 4 When this happens, this FM/NM heterojunction can develop the *spin-Seebeck effect* (SSE).[3,9] The two mixed effects, MD and SSE, will be described in section 6.

(3) For completion, we add that spin also can be transported by the motion of magnetic domains in a sample, although this will not be described any further.

A few more particularities to spin transport need to be mentioned.

The first difference between spin and heat or charge transport results from the fact that while heat and charge are scalars, magnetization and spin-angular momentum are vectors: they point in the direction $\vec{\sigma}$ (a unitary vector) of the spin polarization (so $\vec{S} \parallel \vec{\mu}_m \parallel \vec{\sigma}$). In practice, $\vec{\sigma}$ either is imposed by an external applied magnetic field, or is aligned with the magnetization of a FM sample. In general, $\vec{\sigma}$ is different from the propagation direction of the spin flux $\vec{j}_S$, which is thus formally a tensor. For simplicity, we keep using a vector notation for $\vec{j}_S$, with the arrow denoting its propagation direction.

Second, quantifying spin transport requires developing a technique to measure spin fluxes, a "spin-ammeter" so to speak. The usual method is to evaporate a Pt film on top of a FM sample, and rely on the *inverse spin-Hall effect* (ISHE). We will describe this in detail in section 5.

Third, unlike charge, spin is not conserved; it decays naturally over the scale of nanometers to microns in the solids in which it resides. This is not a problem for the Onsager relations, but it requires the introduction of one additional concept: the spin lifetime $\tau_S$ and the accompanying spin diffusion length $L_S = \sqrt{D\tau_S}$ : they are related by the usual diffusion relation with diffusion constant $D$. The diffusion constant itself depends on whether the spin resides on spin-polarized electrons or in magnons (see sections 3 and 4).

Fourth, the conjugate force for spin transport in the Onsager relations is in principle the Landau-Lifshitz effective field $\vec{H}_{eff}$.[10] The $(\vec{M}, \vec{H}_{eff})$ pair enters Onsager symmetry on par with other thermoelectric quantities. As all thermodynamic potentials, $\vec{H}_{eff}$ is the derivative of the free energy with respect to the magnetization, a formal definition that does not identify the microscopic nature of $\vec{H}_{eff}$. It has contributions from the applied, anisotropy, and exchange magnetic fields. The applied field $\vec{B} = \mu_0 \vec{H}_{ext}$ generates a force $\vec{F} = \nabla(\vec{M} \bullet \vec{B})$ that drives $\vec{j}_S$; anisotropy and exchange fields (explained in section 3) follow the same treatment. To this we add the concept of *spin chemical potential* $\mu_S$ for the spin systems studied here. The magnetic force is then its gradient, $-\nabla\mu_S$. The exact nature of $\mu_S$ will be discussed in detail in the subsequent sections because it is defined differently for spin-polarized electrons, which are fermions, and magnons, which are bosons.

**2. Spin-polarized electrons**

The densities of spin-up and spin-down electrons in metals and semiconductors are labeled $n\uparrow$ and $n\downarrow$, respectively. In non-magnetic metals and semiconductors, and in the absence of spin injection, $n\uparrow = n\downarrow$. Spin polarization can occur in metals and semiconductors by an external magnetic field, by the net magnetic moment that develops in magnetically aligned materials, FMs and ferrimagnets, or by direct spin-injection of carriers of one spin polarization. When the spin relaxation is weak, i.e., in the limit for $\tau_S$, $L_S \rightarrow \infty$, one can approximate FM metals by a two-fluid model: spin-up and spin-down electrons, which use spin-up and spin-down densities $n\uparrow \neq n\downarrow$ as well-defined thermodynamic quantities (and which could be conserved approximately) that enter Onsager reciprocity relations. Transport of charge current then is accompanied by a spin current.

In the two-fluid model, Fig. 2.1, the spin-up and spin-down electron bands are distinct. The electrochemical potential level at equilibrium is the same for all bands, so that the chemical potentials $\mu\uparrow$ and $\mu\downarrow$ for spin-up and spin down electrons, measured vis-à-vis their band edges, are distinct. In the presence of a gradient in these potentials, generated, e.g., by an electric field, the Onsager relations relate the charge current densities in the two fluids via their partial conductivities $\sigma_\uparrow$ and $\sigma_\downarrow$:

$$\begin{bmatrix} j_\uparrow \\ j_\downarrow \end{bmatrix} = \begin{bmatrix} \sigma_\uparrow & 0 \\ 0 & \sigma_\downarrow \end{bmatrix} \begin{bmatrix} -\nabla \mu_\uparrow \\ -\nabla \mu_\downarrow \end{bmatrix} \quad (2.1).$$

We define the charge current $j_C$ and the spin current $j_S$ by:

$$\begin{aligned} j_C &\equiv j_\uparrow + j_\downarrow \\ j_S &\equiv \frac{\hbar}{e}\left(j_\uparrow - j_\downarrow\right) \end{aligned} \quad (2.2),$$

and define the average chemical potential as $\mu$ and the spin chemical potential as $\mu_S$:

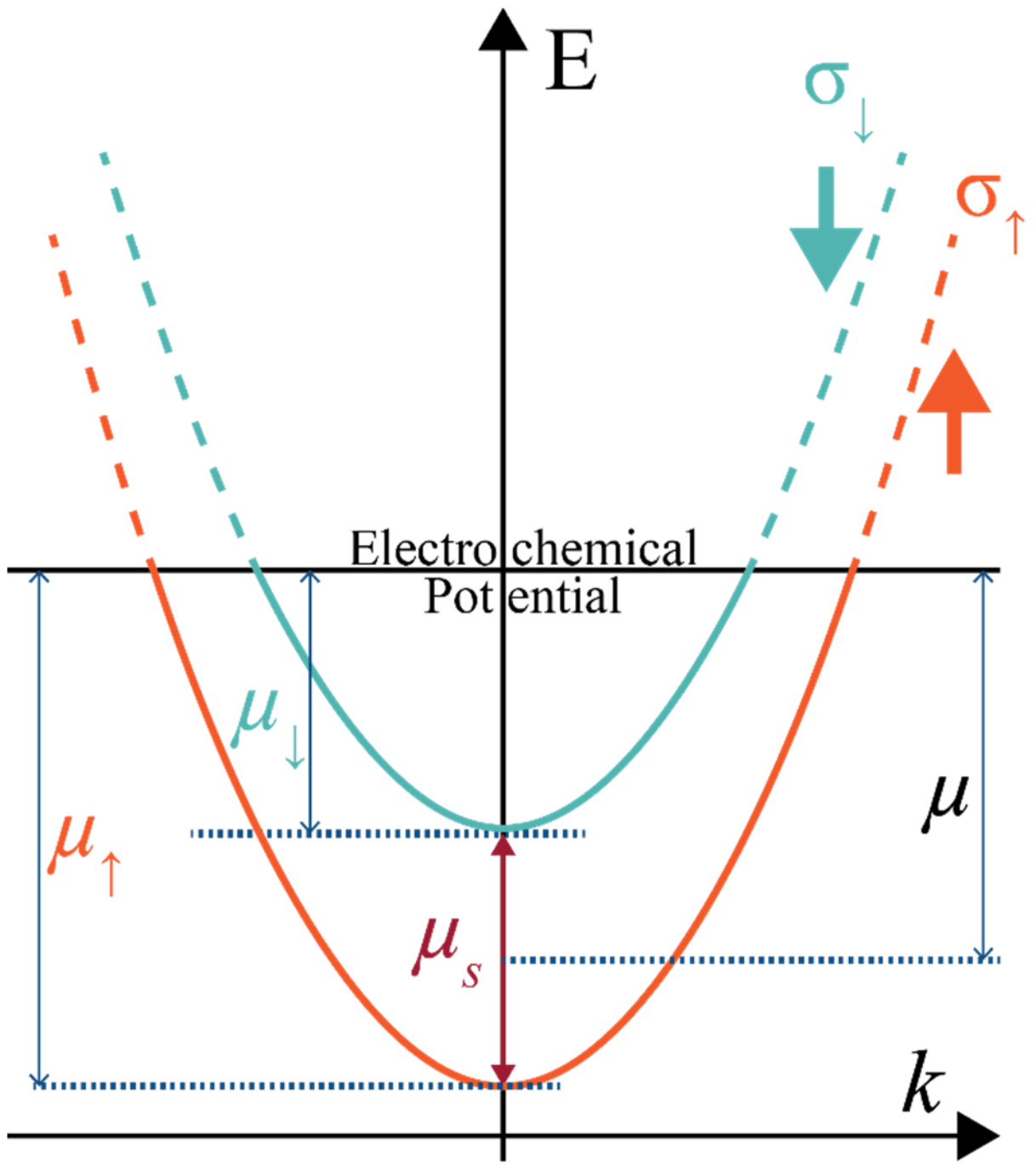

Fig. 2.1 The two-fluid model for conduction by spin-polarized electrons.

$$\begin{aligned} \mu &= \frac{1}{2}\left(\mu_\uparrow + \mu_\downarrow\right) \\ \mu_S &= \left(\mu_\uparrow - \mu_\downarrow\right) \end{aligned} \quad (2.3),$$

and the electrical conductivity $\sigma$ and the spin conductivity $\sigma_S$ as:

$$\begin{aligned} \sigma &= \sigma_\uparrow + \sigma_\downarrow \\ \sigma_S &= \sigma_\uparrow - \sigma_\downarrow \end{aligned} \quad (2.4).$$

Substituting (2)-(4) into (1) gives a new Onsager relation (5) that now relates the charge current to the spin current:

$$\begin{bmatrix} j_C \\ j_S \, e/\hbar \end{bmatrix} = \begin{bmatrix} \sigma & \sigma_S \\ \sigma_S & \sigma \end{bmatrix} \begin{bmatrix} -\nabla \mu \\ \frac{-1}{2} \nabla \mu_S \end{bmatrix} \quad (2.5).$$

The gradient in spin chemical potential can have several physical origins. As explained in the introduction, the rigorous conjugate force for spin transport is the Landau-Lifshitz effective field $H_{eff}$. An applied external magnetic field, or the magnetization in the sample, contribute to $H_{eff}$. Thus, a gradient in either external field or in magnetization exerts a magnetic force, $\vec{F} = \nabla(\vec{M}.\vec{H})$ on the carriers.[11] Another mechanism to generate a $\nabla\mu_S$ is to inject spin-polarized carriers into the metal dynamically: examples of how this can be done are given in section 5.

The effect of spin-flipping electron interactions that limit $\tau_S$, when not so intense as to invalidate the two-fluid model completely, are taken into account by using the drift-diffusion equation.[12] Eq. (2.1) then becomes:

$$j_{S\uparrow} = \sigma_\uparrow \nabla\mu_\uparrow - D\nabla n_\uparrow \quad ; \quad j_S = \sigma_\downarrow \nabla\mu_\downarrow - D\nabla n_\downarrow \qquad (2.6),$$

where $D = {\mu_m k_B T}/{e}$ is the electron diffusion constant and $\mu_m$ is the charge carrier mobility, and the presence of gradients in $n\uparrow$ and $n\downarrow$ is related to the spin-flip transitions that govern both.

Consider a one-dimensional picture (Fig. 2.2) where an accumulation of spins is injected into a metal at the left side ($x$=0), with an initial $n_\uparrow(x=0) - n_\downarrow(x=0) = \Delta n_{\uparrow\downarrow,0}$ accumulation. Over a distance $x$ into the metal, spin-flip transitions between the two populations, with a spin lifetime $\tau_S$ will reduce the amount of spin imbalance $\Delta n_{\uparrow\downarrow}(x) = n_\uparrow - n_\downarrow$; thus, the net magnetization $\delta|\vec{M}| = \mu_B \Delta n_{\uparrow\downarrow}$ also will be reduced. This is determined by the diffusion equation

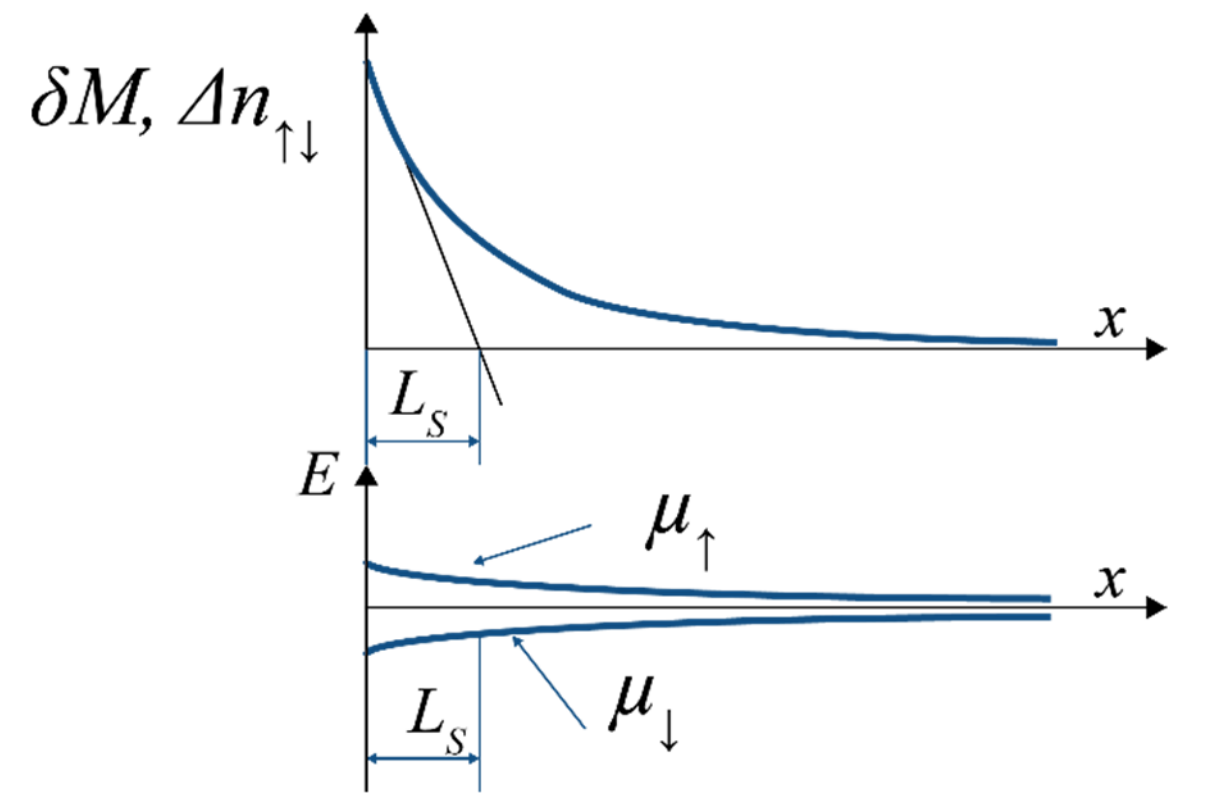

Fig. 2.2 Electron spin diffusion length $L_S$ in the two-fluid model.

$$\Delta n_{\uparrow\downarrow} = \Delta n_{\uparrow\downarrow 0} \exp(-x / L_S) \qquad (2.7),$$

as shown in Fig. 2.2. Here, the diffusion length is $L_S = \sqrt{D\tau_S}$ . Since Fermi-Dirac statistics directly relate the partial charge carrier concentrations to the chemical potentials:

$$n_{\uparrow or \downarrow} = \int_0^\infty \frac{\mathcal{D}(E)dE}{1+\exp(\frac{E-\mu_{\uparrow or \downarrow}}{k_B T})} \qquad (2.8),$$

where $\mathcal{D}(E)$ is the electronic density of states (DOS), $\Delta n_{\uparrow\downarrow}(x) = n_\uparrow - n_\downarrow$ is equivalently represented by a change in $\mu_\uparrow(x)$ and $\mu_\downarrow(x)$, and thus $\mu_S(x)$ as shown in Fig. 2.2.

Adding a temperature gradient to the problem results in a mixed charge-spin-heat Onsager relation:

$$\mathbf{j} \equiv \begin{pmatrix} j_C \\ j_Q \\ j_S \end{pmatrix} = \begin{pmatrix} \sigma & L_{ET} & \sigma_S \\ L_{TE} & \kappa & L_{TM} \\ -\sigma_S & L_{MT} & \sigma \end{pmatrix} \begin{pmatrix} -\nabla\mu_\phi \\ -\nabla T \\ -\nabla\mu_S \end{pmatrix} = \mathbf{LF} \qquad (2.9),$$

with a production of irreversible entropy (spin propagation is dissipative):

$$\dot{S} = \mathbf{j} \cdot \mathbf{F} / T \qquad (2.10).$$

Here, we recognize the classical thermoelectric conductivity $L_{ET}$, which gives rise to the thermopower $\alpha = L_{ET} / \sigma$, but it should be pointed out that this thermopower is driven by spin-polarized carriers; thus, it is a spin-dependent Seebeck coefficient, reviewed by Boona et al.[4] and Vandaele et al.[7] The spin-dependent Peltier conductivity $L_{TE}$ is accompanied by a thermally driven spin flux via the non-zero coefficient $L_{MT}$. The van Wees group have seen the spin-dependent Seebeck[13] and Peltier[14] coefficients experimentally, as have many others.[15,16] In the two-fluid model, the partial thermopowers for spin-up and spin-down electrons, $\alpha_\uparrow = L_{ET\uparrow} / \sigma_\uparrow$ and $\alpha_\downarrow = L_{ET\downarrow} / \sigma_\downarrow$ are given by the Mott formula. The total thermopower is given by the conductivity-weighted average of the partial thermopowers, as is customary for all multi-carrier systems:

$$\alpha = \frac{\alpha_{\uparrow}\sigma_{\uparrow} + \alpha_{\downarrow}\sigma_{\downarrow}}{\sigma_{\uparrow} + \sigma_{\downarrow}}, \quad (2.11).$$

For the purpose of thermoelectric performance, only the total thermopower matters. The two-fluid model has also been used to interpret the Nernst effect in metallic FMs[6].

## 3. Magnons

### *3.1 Ferromagnets*

Consider a FM insulator at 0 K in which all moments reside on the core electrons on unfilled *d*- or *f*-shells of the atoms in the solid. This is the ground state of the system, represented in Fig. 3.1A. The interatomic distance is *a*, and the spins are coupled to each other by the magnetic *exchange energy J*. The $p^{\text{th}}$ atom interacts with its neighbors of index *p*-1 and *p+1*. The ground state energy of the system is:

$$U = -2J\sum_{p=1}^{N} \vec{S}_p \bullet \vec{S}_{p+1} \quad (3.1).$$

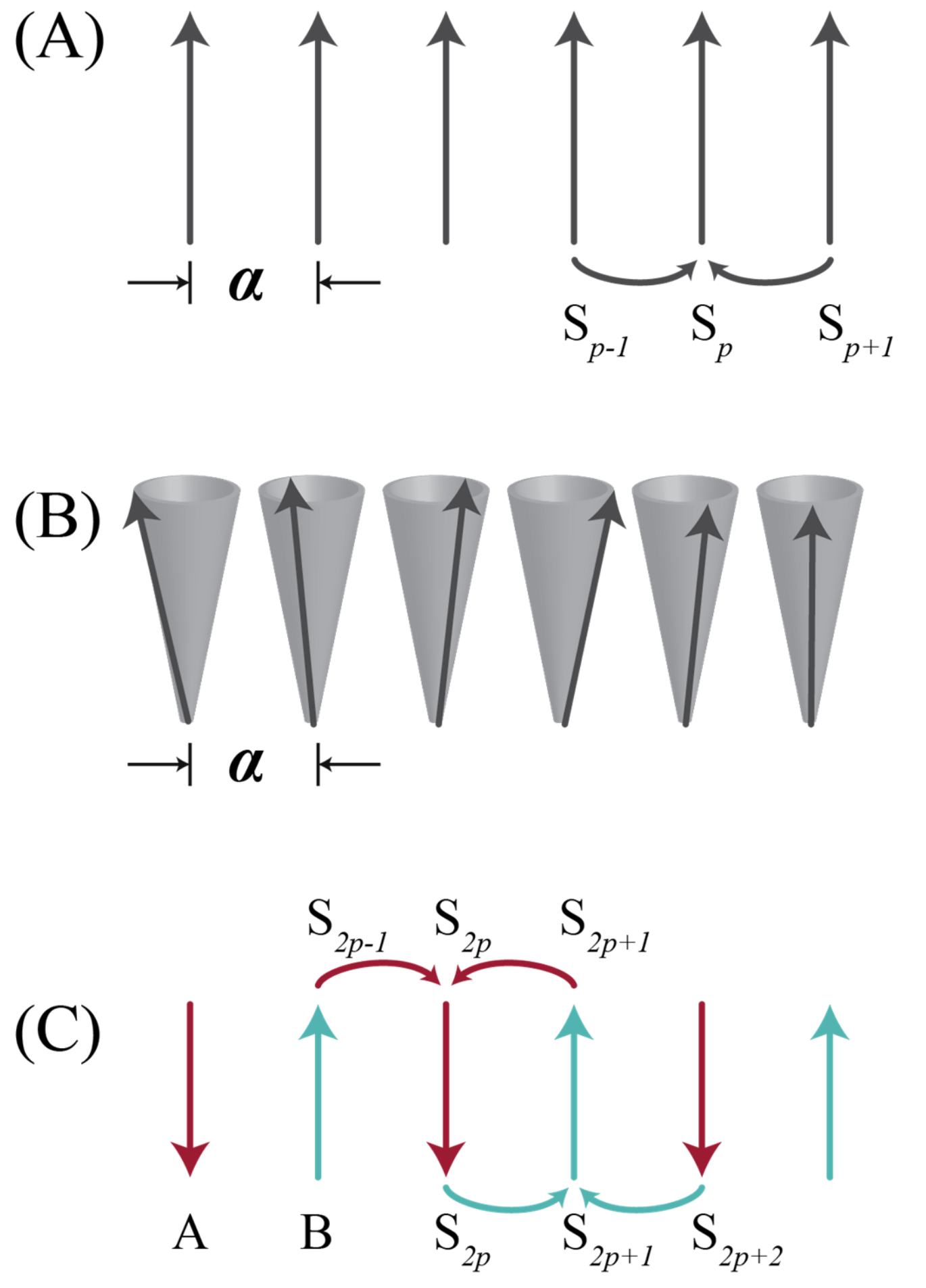

Fig. 3.1 (A) Magnetic moments in a ferromagnet of lattice constant *a* at zero K. Neighboring spins of index p-1, p and p+1 couple. (B) At finite temperature, spin waves develop. (C) Antiferromagnets have two sublattices A and B; two sets of couplings are possible.

At finite temperature, the individual spins do not start flipping arbitrarily through the system, as this would cost too much energy. Instead, all spins share the decrease of magnetization by developing a precession motion, as shown in Fig. 3.1B. The precession motion becomes a wave, called a *magnon*, much like phonons are waves of atomic displacements. The projection of each moment along the direction of magnetization at 0 K, the saturation magnetization $\vec{M}_S(T)$ of the sample at finite temperature *T*, is decreased $.\vec{M}_S(T) < \vec{M}_S(T=0)$. The dynamic magnetization $\vec{m}(\vec{r},T)$ (see Fig. 3.2) is the quantity that will form the wave. To carry the analogy between magnons and phonons further, $|\vec{m}(\vec{r},T)|$ (or the apex cone angle) is, for magnons, the quantity equivalent to the amplitude of the atomic motion for phonons. The phase angle of $\vec{m}(\vec{r},T)$ is equivalent to the phase of the atomic motion in phonon propagation.

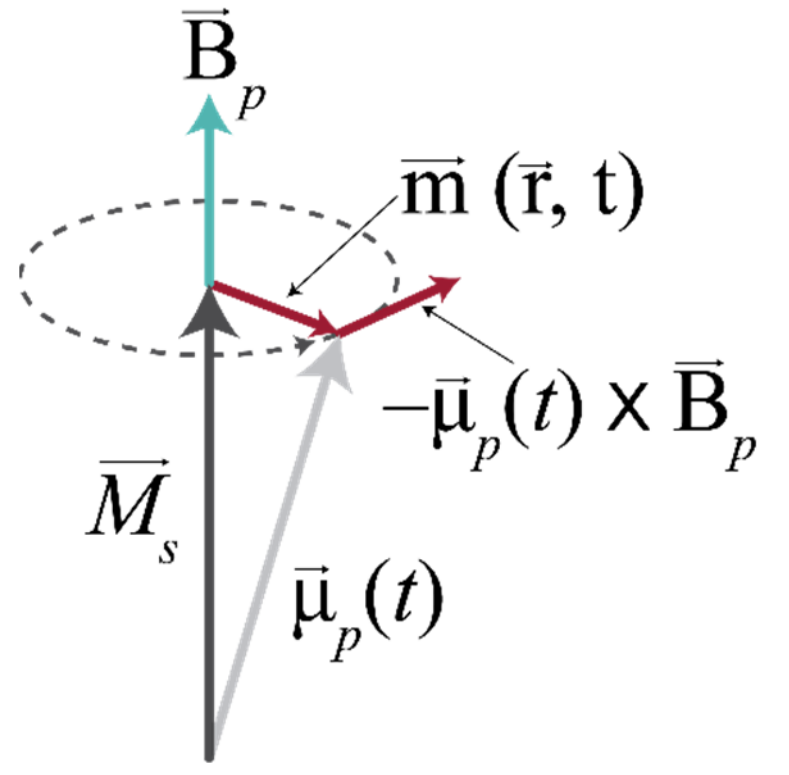

Fig. 3.2 Equation of motion for magnons.

The equation of motion of magnons is different from that of phonons (the ball-and-spring model): the individual moment $\vec{\mu}_p(t)$ of the $p^{\text{th}}$ atom is shown in Fig. 3.2. From interactions with its neighbors (Fig. 3.1) via exchange energy *J*, the effective magnetic induction felt by the $p^{\text{th}}$ atom is:

$$\vec{B}_p = -\left(\frac{2J}{g\mu_B}\right)\left(\vec{S}_{p-1} + \vec{S}_{p+1}\right) \qquad (3.2)$$

This exchange field will generate magnons called *exchange-coupled magnons*. The time-dependence of the moment in the presence of $\vec{B}_p$ is then:[17]

$$\frac{\hbar}{g\mu_B}\frac{d\vec{\mu}_p(t)}{dt} = -\vec{\mu}_p(t)\times\vec{B}_p(t) \qquad (3.3),$$

where the right side of the equation is the torque that drives the precession. Equivalently, one can write:

$$\frac{d\vec{S}_p}{dt} = \left(-\frac{g\mu_B}{\hbar}\right)\left\{\vec{S}_p(t)\times\vec{B}_p(t)\right\} = \left(\frac{2J}{\hbar}\right)\left\{\vec{S}_p\times\vec{S}_{p-1} + \vec{S}_p\times\vec{S}_{p+1}\right\} \qquad (3.4).$$

The solution to Eq. (3.4) in Cartesian coordinates ($z$ being the direction of $\vec{M}_S$) is $S_p^x = u\exp(ipka - \omega t); S_p^y = v\exp(ipka - \omega t)$. The dynamic part of the magnetization, $\vec{m}(\vec{r},t)$ (see Fig. 3.2), with the same periodic boundary conditions as apply to phonon physics, is a propagating wave with a wavevector in k-space and an angular frequency ω:

$$\vec{m}(\vec{r},t) = m_0 \exp\left\{i\left(\vec{k}\cdot\vec{r} - \omega t\right)\right\} \qquad (3.5).$$

The difference between the equations of motion for phonons and magnons in ferromagnets results in a difference between their dispersion relations. Considering only one dimension, the dispersion relation for FM magnons is:

$$\hbar\omega = 4JS(1-\cos ka) \qquad (3.6),$$

which resembles that of electrons in a tight-binding model. At low frequency, Eq. (3.5) gives a Taylor expansion that is parabolic in $k, \hbar\omega \cong \left(2JSa^2\right)k^2$, which more generally is written as $\hbar\omega \cong Da^2k^2$, where $D$ is the magnon stiffness. Here, the magnon stiffness is derived for these exchange-coupled magnons. This quadratic dispersion now looks like that of electrons near the band edge. If we add an external magnetic field $B_{ext}$, it adds a Zeeman energy $g\mu_B B_{ext}$ to the magnon dispersion, which, being independent of $k$, looks like a band gap in the magnon dispersion.

Finally, to all this we add the presence of magnetic anisotropy in the sample, either crystalline or geometrical. The anisotropy energy also can be expressed in terms of an anisotropy field $B_a$, which can simply be added to the external field to form an "effective" field $B_{eff}$. The final magnon dispersion for FMs is then:

$$\hbar\omega = g\mu B_{eff} + Da^2k^2 \qquad (3.7).$$

Experimentally, since neutrons are sensitive to spin, inelastic neutron scattering can be used to map out magnon dispersions as well as phonon dispersions, and the results confirm the calculated dispersion relations quite well, see Fig. 3.3(A).[18] The gap at *k*=0 can be detected by optical techniques, which involve no exchange in *k*-vector. *Ferromagnetic resonance* (FMR) is the classical technique for this that uses microwaves (the gap is typically a few GHz). The microwave absorption shows a maximum when $\omega = g\mu_B(B_a + B_{ext})/\hbar$. The absorption peak can be followed or tuned by applying an external magnetic field $B_{ext}$.

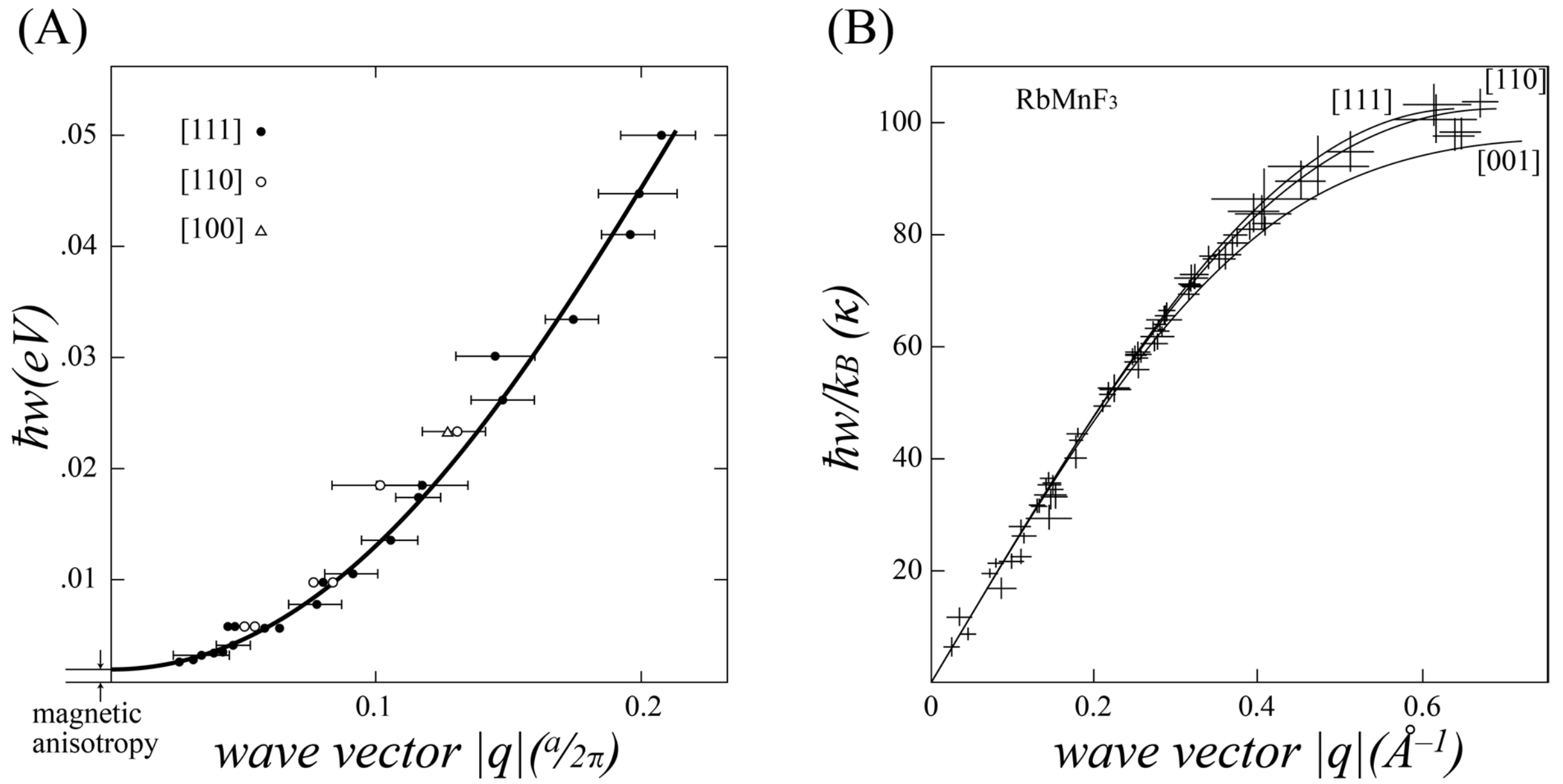

Fig. 3.3 (A) FM exchange-coupled magnon dispersion in cobalt; (B) AFM exchange-coupled magnon dispersion. Redrawn and adapted from (A) Ref. 18 and (B) Ref. 19.

In addition to exchange-coupled magnons, there are *dipole-coupled magnons*, mostly at the surface and at very low energy, which are coupled by dipole interactions. Their dispersion is not necessarily quadratic, and their group velocities are very small and can even be negative ("backward-propagating magnons"). The dipole-coupled magnons are seen by inelastic light scattering techniques such as *Brillouin light scattering*, which involve infinitesimally small k-vectors. Because of their low velocity, dipole-coupled magnons contribute little to transport and will not be considered here further.

*3.2. Antiferromagnets.*

AFMs support magnons as well, but their dispersion is quite different from that of FM magnons. There are many types of AFM ordering in various solids. The simplest consists of the one-dimensional magnetic sublattices A and B, shown in Fig. 3.1 as green and red sublattices. The 3D version of this would be a cubic AFM ordering where each site's spin is the opposite of each of its nearest neighbors ($RbMnF_3$, $LiNiF_3$): this gives a very small magnetic anisotropy. Many other types of AFM ordering exist, such as sheets of FM-ordered planes stacked in an AFM fashion (e.g., MnTe along the <001> direction), or triangular or helical lattices (e.g., one of the phases in many heavy elemental rare-earth metals).[17]

In this paper, we only consider the very simple case of Fig. 3.1, with sublattices A (red) and B (green). There are now two coupled equations of motion like Eq. (3.4), one per sublattice. Assuming that the moments are equal but opposite ($\vec{S}_A = -\vec{S}_B = \vec{S}$), the effective fields for atom index $2p$ of sublattice A and for atom of index $2p+1$ of sublattice B are:

$$\vec{B}^A_{2p} = -\left(\frac{2J}{g\mu_B}\right)\left(\vec{S}_{2p-1} + \vec{S}_{2p+1}\right); \vec{B}^B_{2p+1} = -\left(\frac{2J}{g\mu_B}\right)\left(-\vec{S}_{2p} - \vec{S}_{2p+2}\right) \qquad (3.8).$$

Eqs. (3.3-3.5) are now replaced a system of two equations for indices $2p$ and $2p+1$:

$$\begin{aligned} S^x_{2p} &= u_A \exp[i(2pka - \omega t)]; S^y_{2p} = v_A \exp[i(2pka - \omega t)] \\ S^x_{2p+1} &= u_B \exp[i\left(\left(2p+1\right)ka - \omega t\right)]; S^y_{2p+1} = v_B \exp[i(2\left(2p+1\right)ka - \omega t)] \end{aligned} \qquad (3.9)$$

This system has roots only if $\omega^2 = \left(-4JS/\hbar\right)^2 (1 - \cos^2(ka))$ and the dispersion relation becomes:[17]

$$\hbar\omega = \hbar\omega_{\max} \left|\sin(ka)\right| \qquad (3.10).$$

Interestingly, while the magnon dispersion in a FM solid looks like the dispersion of electrons, magnon dispersion in an AFM looks like the dispersion of phonons. The Taylor expansion at low energy is linear:

$$\hbar\omega = \hbar\omega_{\max} ka \qquad (3.11).$$

As for FM magnons, magnetic anisotropy adds a Zeeman term, which looks like an energy gap. The magnon dispersion relation measured by neutron scattering on $RbMnF_3$, which has negligible anisotropy, is shown in Fig. 3.3(B)[19] and follows Eq. (3.10) perfectly.

*3.3 Equilibrium thermodynamic properties of magnons.*

The DOS $\mathscr{D}(E)$ of FM and AFM magnons is calculated from the dispersion relations as for all other quasiparticles, and is shown in Fig. 3.4. Given the similarity between FM magnons and electrons, it is not surprising that their DOS follows a $\sqrt{E}$ or $\sqrt{\omega}$ law, with an offset that is the gap Eq. (3.7). In particular, this gap can be changed by applying an external magnetic field. Likewise, given the similarity of the AFM magnon dispersion to that of phonons, the AFM DOS follows a $E^2$or $\omega^2$ law.

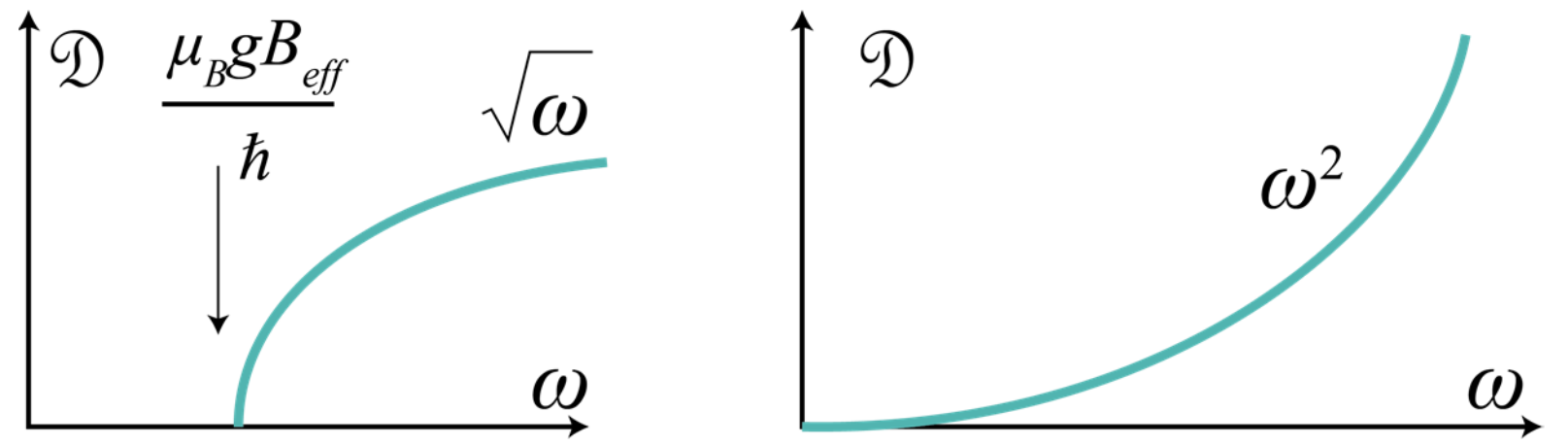

Fig. 3.4 Magnon DOS, (A) in a FM; (B) in an AFM.

Like phonons and electrons, magnons carry heat and entropy, in an amount $k_B$ per particle. Thus, there is a magnon specific heat $C_m$, calculated like with all other quasiparticles as the temperature derivative of the internal energy $U$ of the system. $U$ is obtained by integrating the energy per quasiparticle over the ensemble of particles, itself obtained by using the appropriate DOS and statistical distribution function.

For FM magnons, this results in a $C_m \propto T^{1.5}$ law at low temperature (see Fig. 3.5[4]). The existence of a field-dependent gap in the dispersion offers a way to separate $C_m$ from the other contributions to the specific heat $C$: $C_m$ can be frozen out by applying a high magnetic field[4]. At zero field, $C=C_P+C_e+C_m$, in yttrium iron garnet (YIG), an insulating FM, $C_e$ =0). As the applied field is increased, the magnon DOS shifts to a higher energy, and no magnons contribute to $C_m$ at low $T$. As a result, at the highest field, only the phonon specific heat is measured. Therefore, by taking the difference between $C(B_{ext}=0$ Oe) and $C(B_{ext}=7$ kOe), one can isolate $C_m$.

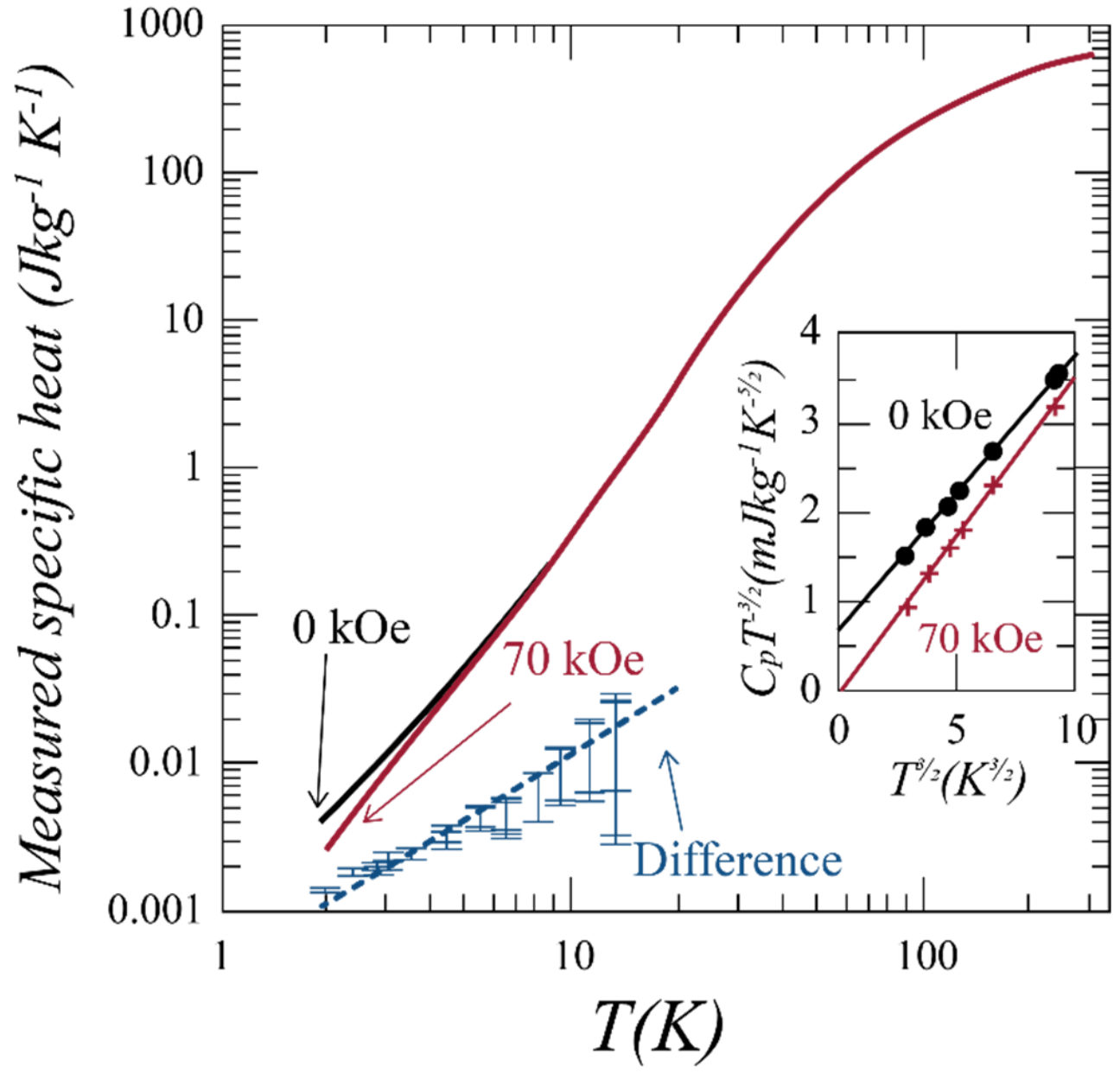

Fig. 3.5 Specific heat in yttrium iron garnet (YIG), an insulating FM garnet. Applying an external magnetic field isolates the phonon contribution. The magnon contribution can then be obtained from the difference between measurements at zero field and at 70 kOe.[4]

The specific heat of AFM magnons at temperatures far below the ordering or Néel temperature ($T_N$) is congruent to that of phonons because the energy dependence of the dispersion relation and the DOS are congruent for both quasiparticles, and their statistical distribution functions are the same. At $T << T_N$, $C_m$ follows a Debye-like law with $\hbar\omega_{\max}/k_B$ as a magnon cutoff temperature. At low temperature, $C_m \propto T^3$ and it is practically impossible to separate $C_m$ experimentally from the phonon contribution. The specific heat of the AFM MnTe is shown in Fig. 3.6[8]. An electronic contribution is observed in this heavily doped sample, but the magnon contribution cannot be resolved from the phonon contribution at low temperature. However, above 150 K and especially near the ordering temperature ($T_N$ =305 K), an additional heat appears over the behavior expected from phonons. Given that the Debye temperature for MnTe (217 K) is much lower than $T_N$, $C_p$ has nearly reached its Dulong-Petit value at $T_N$. Thus, values for $C_p$ can be obtained with reasonable accuracy from a Debye model fit (the full line in Fig. 3.6). Subtracting this from the data gives values for $C_m$ in the 150 – 350 K range: they follow a $T^3$ law, as predicted, except very near and above $T_N$. It is not surprising that an excess heat should appear near $T_N$ because the behavior of magnetic lattices near their melting points is governed by the physics of critical phenomena, and not magnon physics.

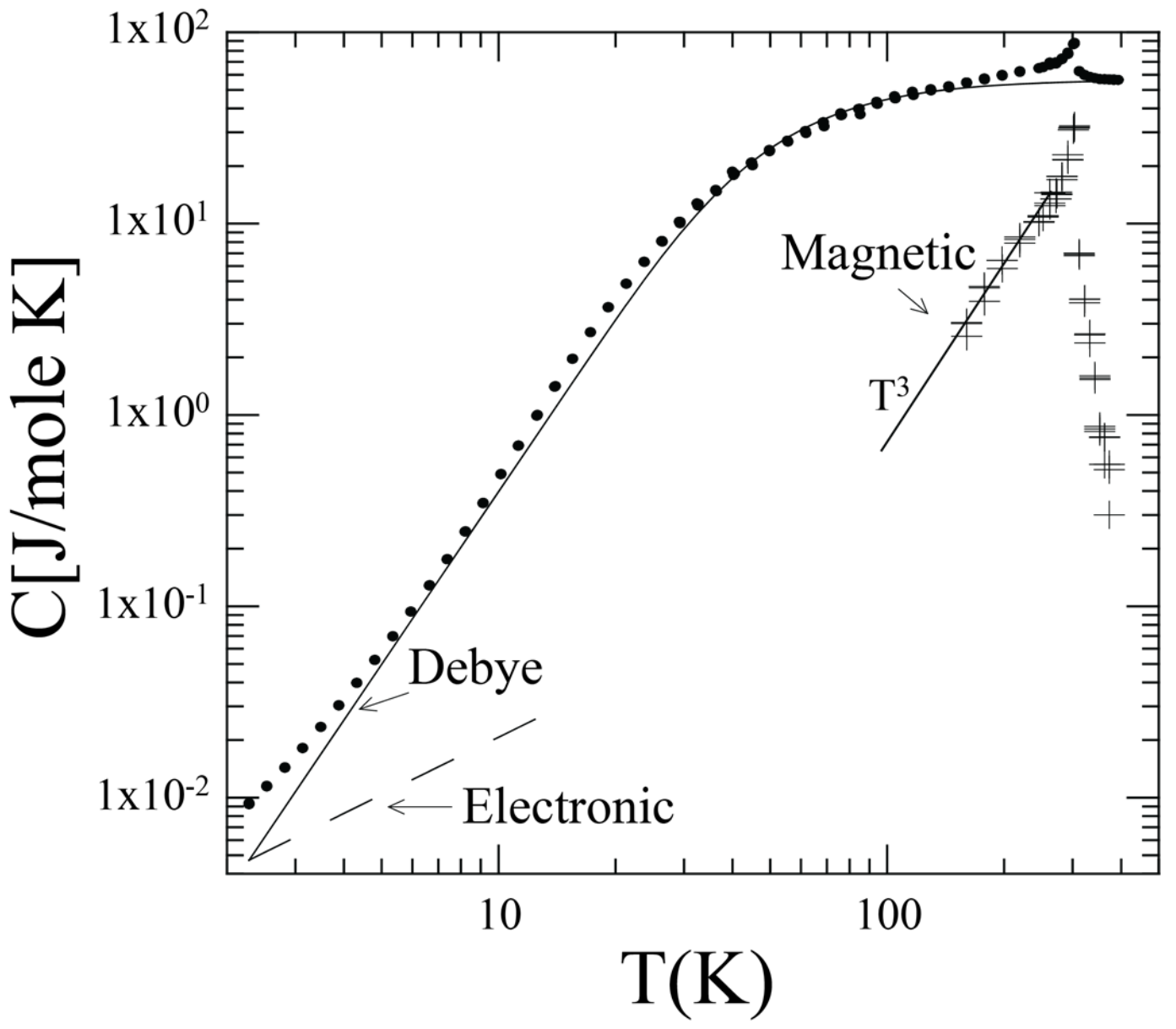

Fig. 3.6 Specific heat of the AFM MnTe. The magnon contribution is isolated near the ordering temperature $T_N$ = 305 K.[4]

*3.4 Magnon thermal transport*

The approach used for the specific heat applies to the magnon thermal conductivity $\kappa_m$ as well, and one can use the kinetic formula:

$$\kappa_m = \frac{1}{3} C_m v_m \ell_m \tag{3.11}$$

for each magnon mode and frequency. In FMs, where one can freeze out the contribution of $C_m$ by applying a magnetic field that opens a Zeeman-energy gap in the dispersion, the same technique can be applied to freeze out $\kappa_m$. This was done for YIG by Boona et al.[4], and the results are shown in Fig. 3.7. The magnetic-field dependence of the total thermal conductivity $\kappa = \kappa_p + \kappa_m(B_{ext})$ is given as function of $T$ and the applied magnetic field. Here $\kappa_p$ is the phonon thermal conductivity. The freeze-out of $\kappa_m$ is visible in its field dependence, which shows a saturation at low temperature. Assuming that this saturation value is $\kappa_p$, the value for $\kappa_m(T)$ can then be obtained by taking the difference as $\kappa_m = \kappa(0\ T) - \kappa(7\ T)$, as shown.

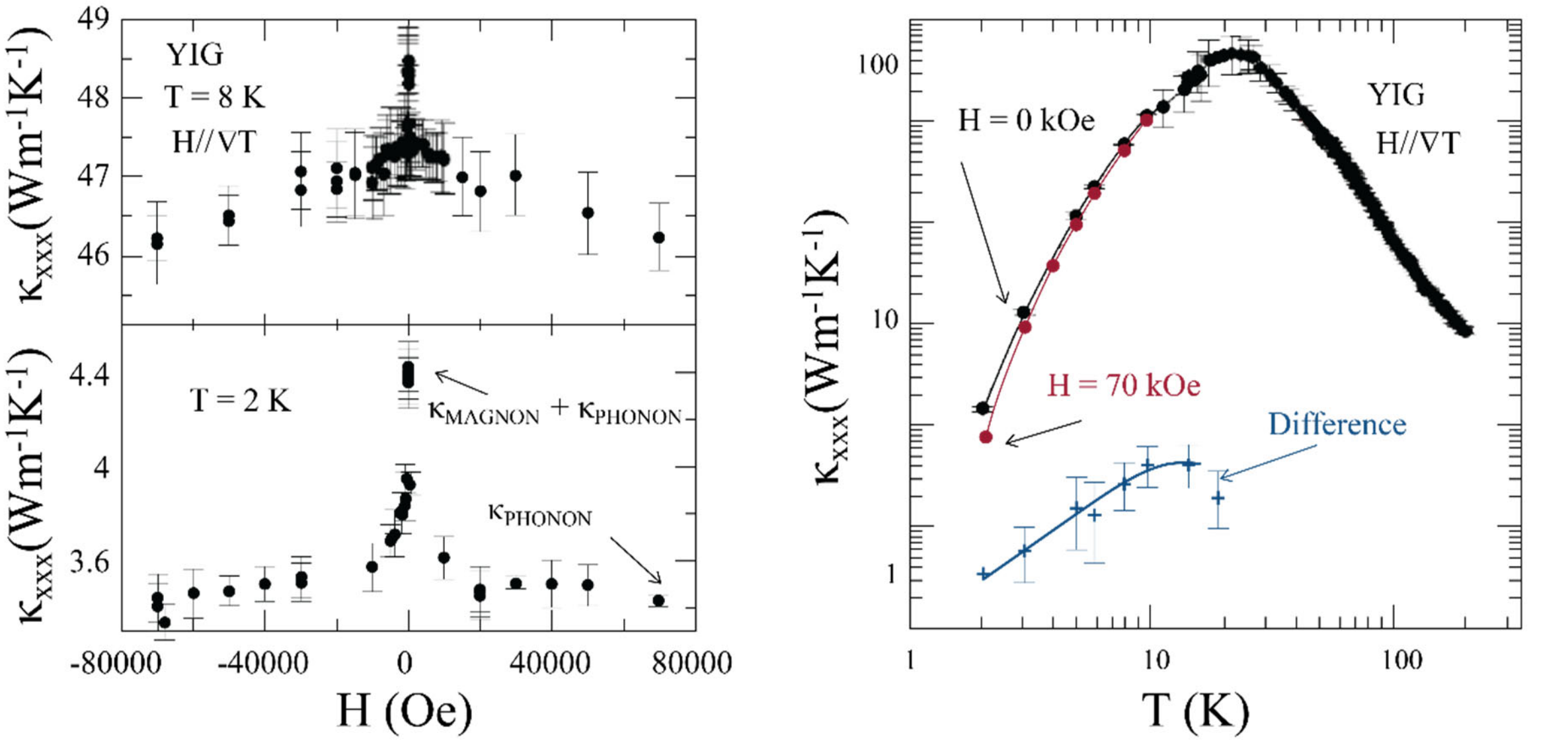

Fig. 3.7 Magnon thermal conductivity of YIG, an electrically insulating FM, in an applied external magnetic field.[4]

Magnons can scatter phonons as well as carry heat; when this is their dominant effect, they lower $\kappa_p$ often to the extent that their contribution results in reducing $\kappa$ as opposed to enhancing it. An example is shown in Fig. 3.8,[20] where the thermal conductivity of non-magnetic $CaF_2$ and AFM $MnF_2$, which have the same crystal and phonon structure, are compared. Below $T_N$, $MnF_2$ has a much lower conductivity, hinting at magnon scattering of phonons. This is more pronounced when non-magnetic $ZnF_2$ is compared to the AFM $CoF_2$, where scattering of phonons by the magnons at $T_N$ is particularly intense, reminiscent of the very large, excess magnetic specific heat near $T_N$ and already shown in Fig. 3.6.

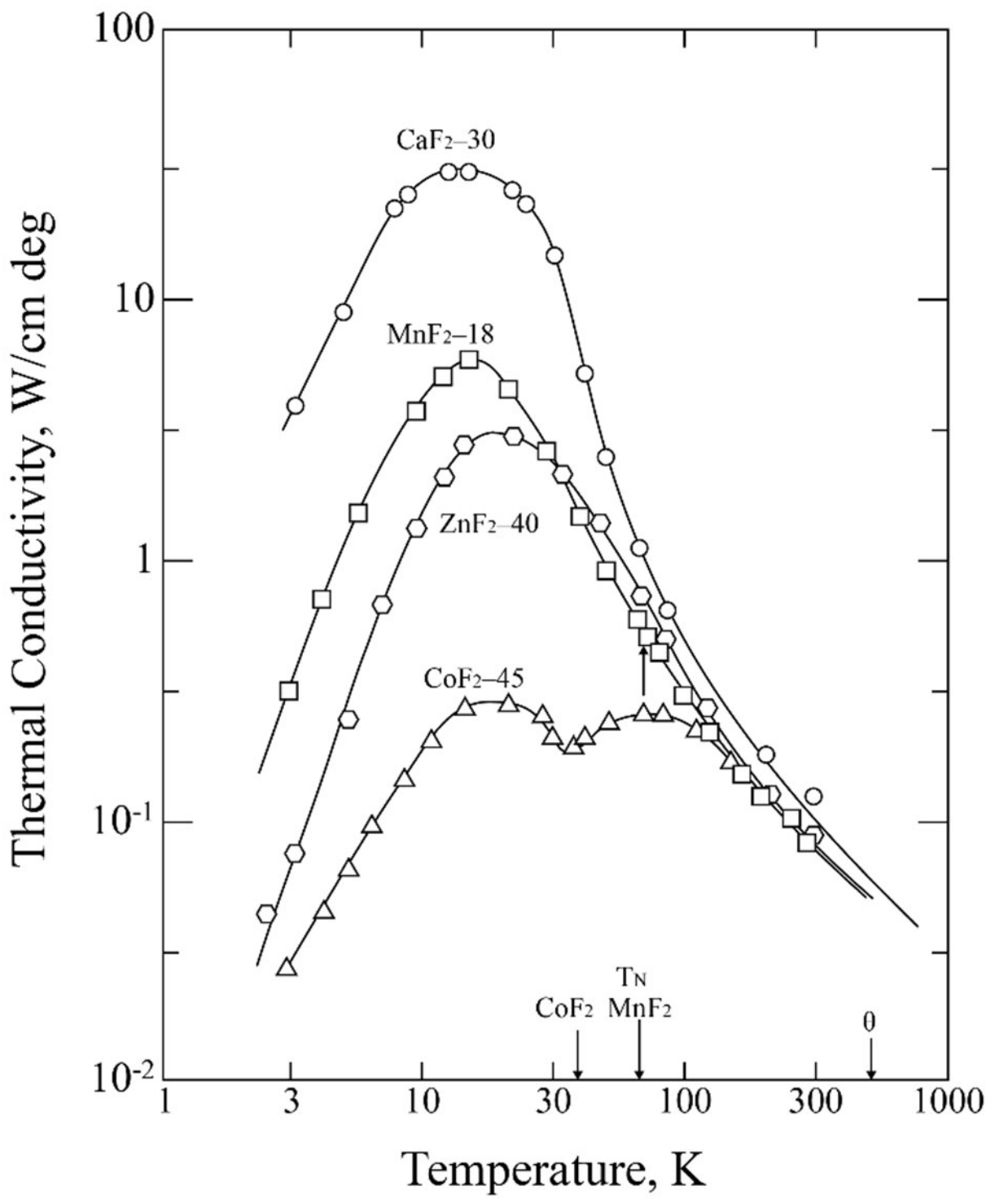

Fig. 3.8 Magnon scattering of phonons. From Ref. 20.

Since all propagating excitations contribute to $\kappa$, its measurement can provide information about transport in insulating FMs and AFMs that is not accessible experimentally otherwise. In such materials, entropy and spin are the only two extensive quantities whose flux can be measured. This paper considers only the most elemental versions of spin waves, but magnon physics is very rich. Magnons can develop esoteric topological properties and, in principle, develop topologically protected transport. One particularly interesting form of magnetization propagation are spinons in quantum spin liquids (QSLs), collective modes that appear in frustrated magnets and are not bosons, but fermions. Gapless fermionic spinons are expected to have $\kappa \propto T^1$,[21] as opposed to the $T^3$ for $\kappa_p$; this temperature dependence is considered the fingerprint of fermionic particles.

*3.5 Thermal Hall effect*

If one breaks time-reversal symmetry on such topological magnon systems, typically by adding a magnetic field, topologically non-trivial spin structures can generate a thermal Hall effect $\kappa_{xy}$. The measurement of $\kappa_{xy}$[22,23] can provide definitive evidence for chiral topological phases that host a gapped bulk, together with gapless chiral-edge spin excitations,[24,25] e.g., in chiral spin liquids and fractional quantum Hall effects[26,27]. When edge states dominate spin transport, the conduction of heat becomes more pronounced along the edge that allows conduction from hot to cold than along the other, giving rise to a transverse temperature gradient and thus, a thermal Hall effect. This signature feature has been observed in α-$RuCl_3$.[23] Hirschberger et al. report a thermal Hall effect in a frustrated quantum magnet[28] and Kagome magnet[29]. Measurements of $\kappa_{xy}$ are particularly difficult because the copper used in most measurement instrumentation, and in heat spreaders and heat sinks, has an electronic thermal Hall effect (the Righi-Leduc effect) that impose spurious transverse gradients on the samples: instruments have to be redesigned with care with this in mind.

*3.5. Thermally driven magnon spin currents*

In the simplest possible picture, the number flux of magnons, $j_{\#}$ is related directly to the magnon heat flux, as each magnon carries $k_B T$ of heat and the spin flux $j_S$ or flux of magnetization $j_m$, as illustrated in Fig. 3.9.

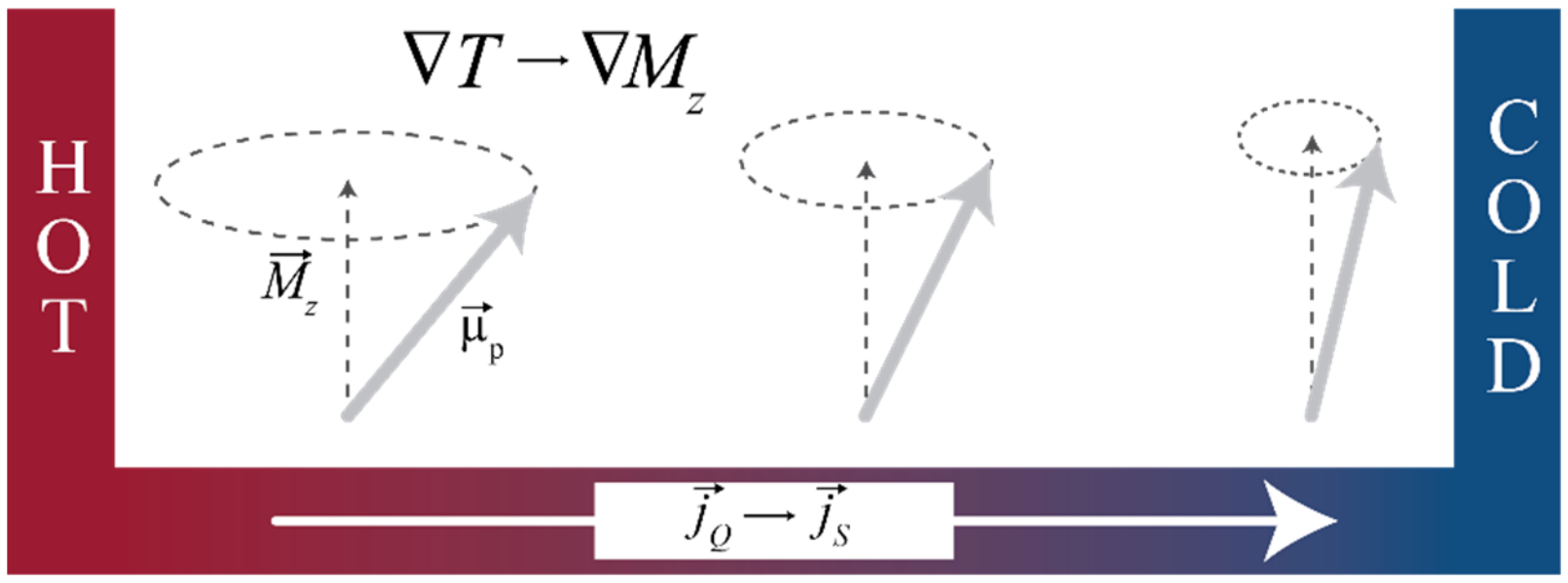

Fig. 3.9 Magnon spin current and magnon heat current.

$$\begin{aligned} j_Q &= -\kappa_m \nabla T \\ j_Q &= k_B T . j_{\#} \\ j_S &= \hbar . j_{\#} \\ j_m &= g \mu_B . j_{\#} \end{aligned} \qquad (3.12).$$

As magnons move from hot to cold, the local saturation magnetization $\vec{M}_S$ (the projection of the atomic moment onto the vertical axis) decreases, which amounts to a transport of magnetization from hot to cold. This is expressed by Eq. (3.12).

*3.7 Spin chemical potential for magnons*

The theoretical concept of the existence of a spin chemical potential for magnons has been proposed recently. It is long accepted that magnons at thermal equilibrium obey Bose statistics with no chemical potential. However, two recent experiments[30,31] have demonstrated that in the presence of a spin current injected by a source external to the sample, a magnon gas can be described as being in quasi-thermodynamic equilibrium with Bose statistics and both a temperature and a spin chemical potential $\mu_S$. The external source of spin current can be FMR pumping or the ISHE in an adjacent layer (see section 5.3). Just like the electronic $\mu_S$ Eq. 2.3, the magnon $\mu_S$ is useful to characterize how the spins residing on magnons diffuse. If an external source pumps an excess of $\Delta n_S$ spins into the magnon system at the surface of a sample (Fig. 3.10), the excess can be described at each point in the sample by:

$$\Delta n_S(x) = \int_0^{\infty} \frac{\mathcal{D}(E)dE}{1-\exp(\frac{E-\mu_S(x)}{k_B T})} \quad (3.13).$$

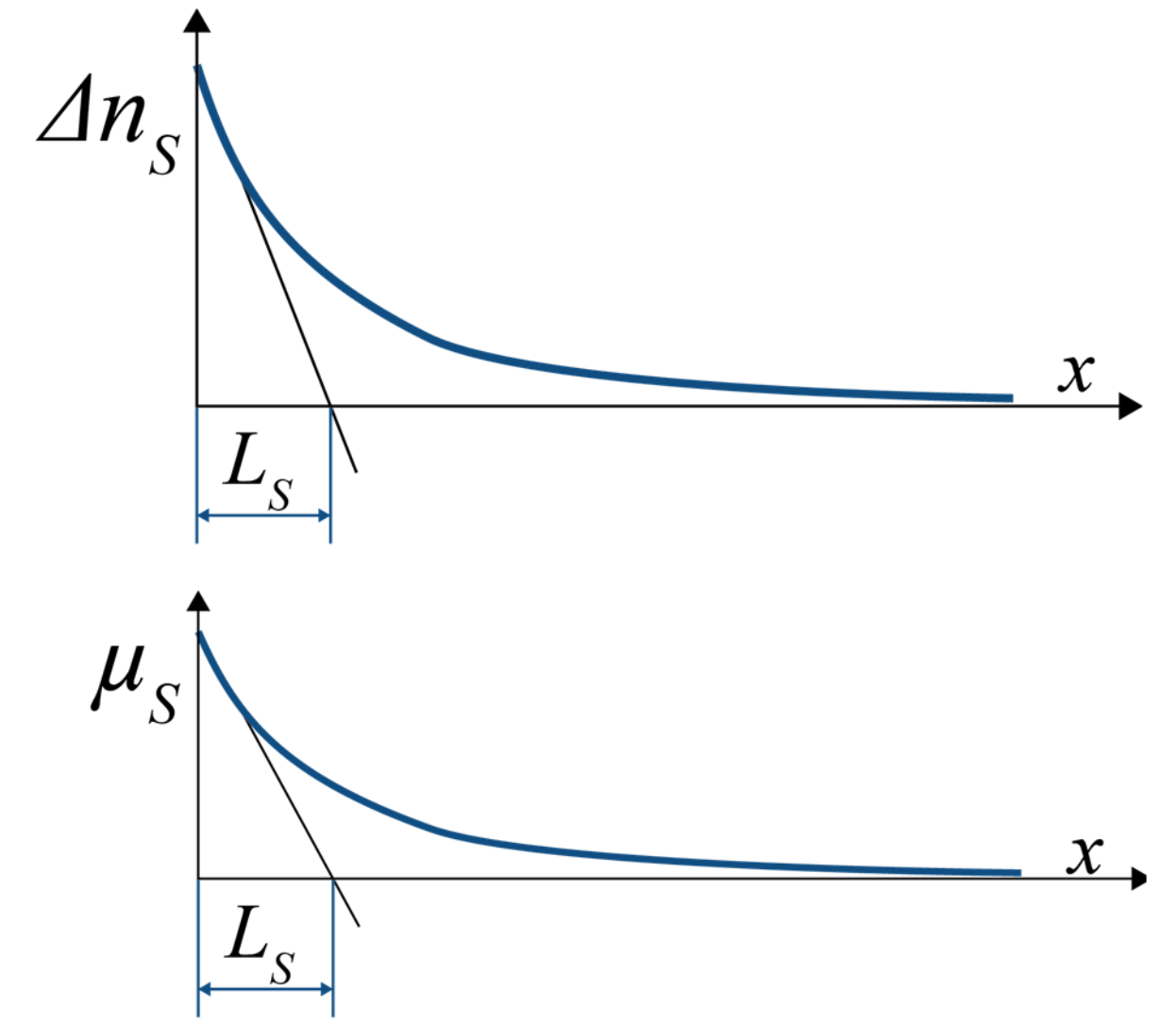

Fig. 3.10 Spin chemical potential for magnons.

The decay of the excess spins is governed by a characteristic diffusion length $L_S = \sqrt{D\tau_S}$ where $D$ is the thermal diffusion constant of magnons and $\tau_S$ is the spin lifetime limited by spin-flip transitions. For magnonic systems, electron-magnon interactions typically are the main source of spin-flip transitions, so the metallic FMs have much shorter spin lifetimes and diffusion lengths than FM insulators, where $L_S$ can reach tens of micrometers. Further, in FM insulators $L_S >> \ell_m$, the mean free path, which correspond to interactions that change the moment of the magnon. Thus, it is only very rarely that a magnon scattering event flips its spin. The concept of spin chemical potential is quite useful in developing magnon transport theories in ferromagnetic insulators.[32] The concept of magnon chemical potential recently has been extended to apply to the sublattices of AFM insulators.[33]

*3.8 Magnonic thermopower*

Gradients in the spin chemical potential can be treated as conjugate forces for spin transport in Onsager relations. By analogy with the thermoelectric Onsager relations, mixed linear thermo-spin Onsager relations[7] connect magnonic spin and heat currents:

$$\begin{pmatrix} j_S \\ j_Q \end{pmatrix} = \begin{pmatrix} \sigma_S & \zeta \\ \pi_m & \kappa_m \end{pmatrix} \begin{pmatrix} -\nabla \mu_S \\ -\nabla T \end{pmatrix} \quad (3.14)$$

where $\kappa_m$ is the magnon thermal conductivity described above. The spin conductivity $\sigma_s$ describes the strength of the spin flux driven by a gradient $\nabla\mu_S$ in the spin chemical potential. New is $\zeta$, the *thermomagnonic conductivity*, the equivalent for magnons of the coefficient $L_{ET}$, the thermoelectric conductivity, used for free electrons. New also is the ratio $\alpha_m \equiv -\zeta/\sigma_S$, the *magnonic thermopower* defined by taking Eq. (3.14) under spin-open-circuit conditions (setting $j_S=0$), where we allow for spin accumulation to occur:

$$\alpha_m \equiv -\frac{\nabla\mu_m}{\nabla T} = -\frac{\zeta}{\sigma_S} \tag{3.15}$$

Both $\alpha_m$ and $\zeta$ have Onsager reciprocals, the magnonic Peltier coefficient and the magnon Peltier conductivity $\pi_m$; these contains essentially the same physics as $\zeta$.

To obtain a microscopic expression for the magnonic thermopower, one can treat the magnon gas as an ideal gas of free particles with internal energy density $U\,(T,\,\mu_S)$. In the presence of a thermal gradient $\nabla T$, an inhomogeneous distribution of magnons arises through the system, which can be expressed in terms of a non-vanishing spin chemical-potential gradient $\nabla\mu_S$. The total gradient in the internal energy is now:

$$\nabla U = C_m \nabla T + n_S \nabla\mu_S \qquad (3.16).$$

This exerts a force $\vec{F}$ that drives the magnon flow. The force is the magnon pressure *P,* in an ideal gas $P = 2/3U$, on a unit surface. Newton's second law, applied to a volume $\delta V$ of the magnon gas, gives:

$$n_S M\,\delta V \frac{d\vec{v}_m}{dt} = \delta\vec{F} \qquad (3.17),$$

where $M$ is the magnon mass and $\vec{v}_m$ its drift velocity. Combining Eqs. (3.16) and (3.17), and then dividing by $\delta V$ gives:

$$\frac{d\vec{v}_m}{dt} = -\frac{2}{3n_S M}\frac{\partial U}{\partial T}\nabla T - \frac{2}{3n_S M}\frac{\partial U}{\partial \mu}\nabla \mu_S = -\frac{2}{3n_S M}C_m \nabla T - \frac{2}{3M}\nabla \mu_S \quad (3.18).$$

The condition $j_s = 0$ means that $\frac{d\vec{v}_m}{dt} = 0$. Eq. (3.17) now gives the magnonic thermopower as:

$$\alpha_m = -\frac{\nabla \mu}{\nabla T} = \frac{C_m}{n_m} \quad (3.19).$$

It is important to note that unlike the electronic thermopower that decreases with increasing concentrations of charge carriers, the magnonic thermopower is the specific heat per spin carrier and thus does not decrease with the density of spin carriers, but rather is a constant of the order of $k_B$. This observation is a consequence of the fact that magnons are bosons, while electrons, being fermions, are subject to the Mott relation between thermopower and density.

**4. Spin-Hall and Anomalous Hall effects**

One of the most useful tools in spin transport technology is the ability to generate and detect spin currents by means of the spin Hall effect (SHE) and the inverse spin Hall effect (ISHE). Both spin-orbit coupling (SOC) in all materials and the presence of permanent magnetic moments in FMs give rise to these effects. In FMs, the SHE is also closely related to the anomalous Hall effect (AHE), which was discovered by E. Hall himself.[34] A schematic representation of the definitions of AHE, SHE and ISHE in FM's, and of SHE and ISHE in NMs with strong SOCs, is given in Fig. 4.1. Excellent reviews exist on this topic.[35] The equivalent thermal effect, the Spin Nernst (SNE), Anomalous Nernst (ANE), and planar Nernst (PNE) effects are reviewed in Boona et al.[36]

*4.1 AHE, SHE, and ISHE in ferromagnetic metals.*

Phenomenologically, in FM conductors, the Hall resistivity, measured in the geometry Fig. 4.1A, takes the form $\rho_{xy} = R_H H_z + \rho'_{xy}$, where $\rho_{xy}$ is the measured Hall resistivity, $R_H$ is the ordinary Hall coefficient, $H_z$ is the applied field and $\rho'_{xy}$ is the anomalous contribution. The term $\rho'_{xy}$ generally is defined as $\rho'_{xy} = 4\pi R_{AH} M$, where $R_{AH}$ is the anomalous Hall coefficient and $M$ is the magnetization. A common misunderstanding is to regard the AHE as being simply the ordinary

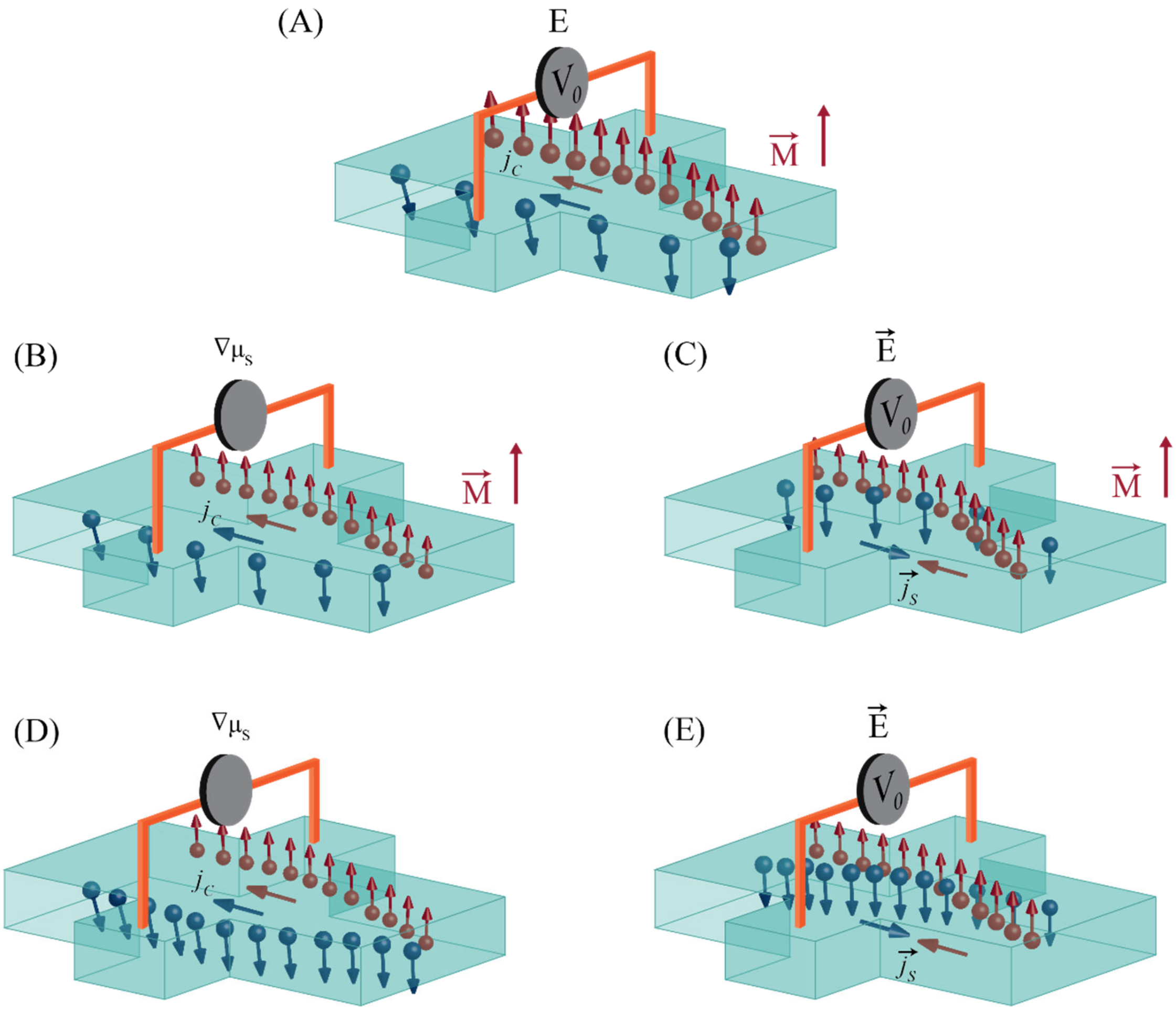

Fig. 4.1 Spin-Hall (SHE), Inverse Spin-Hall (ISHE) and Anomalous Hall (AHE) effects. The directions of the spin polarization, of the spin or charge current propagation, and of the electric field or gradient in spin chemical potential are orthonormal to each other. (A) AHE in FM with magnetization $\vec{M}$ : a charge current $\vec{j}_C$ gives a Hall field $\vec{E}$. (B) SHE in FM with magnetization $\vec{M}$ : a charge current $\vec{j}_C$ gives a spin polarization $\nabla\mu_S$. (C) ISHE in FM with magnetization $\vec{M}$ : a spin current $\vec{j}_S$ gives a gives a Hall field $\vec{E}$. (D) SHE in NM with strong SOC: a charge current $\vec{j}_C$ gives a spin polarization $\nabla\mu_S$. (E) ISHE in NM: spin current $\vec{j}_S$ gives Hall field $\vec{E}$. Expanded from Ref. 35.

Hall coefficient corrected for the real magnetization in the sample; quite to the contrary, $R_{AH} \neq R_H$ and both can even have the opposite signs.

Various mechanisms may produce $R_{AH}$ in FMs, although even after over a century of research, the situation is not always clear because many of these mechanisms are extrinsic and depend on the defect chemistry in the FM. Generally, $R_{AH}$ depends on the material type, temperature, and strength of the applied field.[37,38,39] The approaches are inspired by the two-fluid model Eq. (2.1), to which is added the concept of differential scattering of the spin-up and spin-down electrons. In the first mechanism, *skew scattering*,[40,41] the differential scattering cross-section of the charge carriers that interact with localized impurity states is asymmetric with respect to the carrier spin state. In the second, *side-jump,* mechanism[42], the wave functions of the free electrons are distorted locally during impurity scattering events, as a result of spin-orbit interactions. This causes a spin-dependent offset in the final trajectories of the scattered electrons. Third, the *anomalous velocity* due to the presence of a Berry phase has been invoked as a source of AHE.[43,44,45,46]

The AHE gives rise to several other transverse effects in FM conductors. First is the ANE, which is related to the energy dependence of the AHE by the Mott relation, which holds for metals for transverse thermoelectric coefficients as it does for direct ones:

$$\alpha_{xy} = \frac{\pi^2}{3}\frac{k_B}{e}\frac{k_B T}{\rho_{xy}}\frac{d\rho_{xy}(E)}{dE} \quad (4.1).$$

The second comes from the direct relation between AHE and SHE, illustrated in Fig. 4.1. Again, based on the transformation of variables in Eq. (2.2), the SHE is related to the AHE simply by the fact that the SHE considers the spin accumulation, whereas the AHE considers the charge accumulation that accompany the same effect. The SHE trans-resistance ($-\nabla\mu_S / \left|\vec{j}_C\right|$) is then congruent to the AHE trans-resistance ($|\vec{E}|/|\vec{j}_C|$), multiplied by the appropriate constants. The ISHE in FM metals is represented in Fig. 4.1C. It is the Onsager reciprocal of the SHE: if instead of injecting a charge current and measuring a spin accumulation (the SHE), one injects a spin current, then one must observe a charge accumulation, the ISHE.

*4.2 Spin-orbit Coupling*

SHE and ISHE do not require a spin imbalance to pre-exist in a conductor at thermodynamic equilibrium, and also can be induced by transport in non-magnetic conductors (NMs) with an equal number of spin-up and spin-down carriers, see Fig. 4.1 D and E. It is, in principle, ubiquitous in electrically conducting solids, but really observed only in those where spin-orbit coupling (SOC) is important. The SHE mechanism in NMs is understood much better than the AHE mechanisms in FMs because SOC is intrinsic and not very sensitive to the defect chemistry of the samples. SOC also can be predicted with reasonable accuracy from the band structure or calculated by Density Functional Theory (DFT). The original theoretical idea was published in 1971,[47] but a clean experimental observation of this SHE had to wait for third of a century.[48]

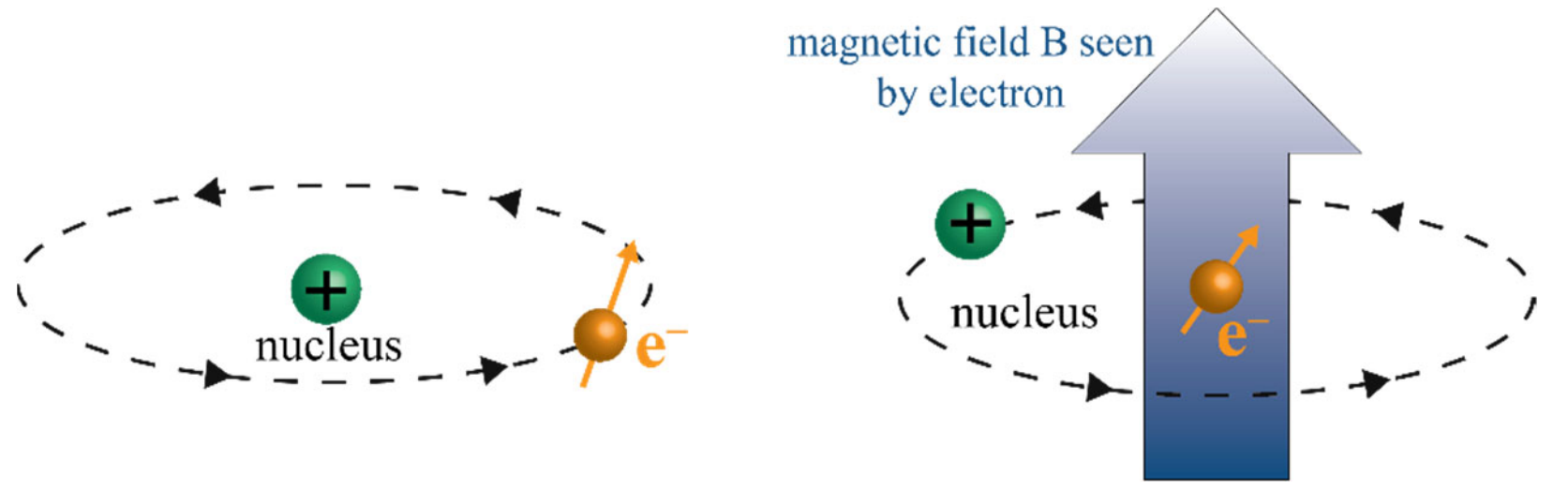

Fig. 4.2 Interaction between the electron spin and the orbital magnetic field around the nucleus of an isolated atom.

SOCs rely on the interactions between orbital magnetic moments and electron spins. Consider first an electron interacting with a single atom in Fig. 4.2. The electron spin interacts with the moment $\vec{L} = \vec{r} \times m\vec{v}$ that arises from the orbital motion of the core electrons around the nucleus (Fig. 4.2). Here, $\vec{r}$ is the radius of the orbit and $\vec{v}$ the electron orbital velocity. The motion gives rise to an orbital magnetic field $\vec{B}_{HO} \propto \vec{L}$, which in turn increases the energy of the electron by a Zeeman term $\propto g\mu_B\vec{\sigma}.\vec{B}_{HO}$. This additional energy means that a term must be added to the Hamilitonian of that electron, $H_{SOC} = \lambda\vec{\sigma}.\vec{L}$, where $\lambda$ is the proportionality constant.

The band structure of solids reflects the equations of motion of electrons as influenced by interactions between electrons and the collective presence of all atoms in the solid (Fig. 4.3). These interactions first take the form of Coulombic interactions between the electron charge and the periodic potential wells *V(r)* that represent the charged atomic nuclei in the solid, screened by the charges on the core electrons (Fig. 4.3 top). The Hamiltonian then has a kinetic energy term and a potential term: $H = \hbar^2 k^2 / 2m + V(r)$. The second contribution comes from the electromagnetic interaction of the spin $\vec{\sigma}$ of the electron with the orbital magnetic fields of the nuclei in the solid (Fig. 4.3 middle). The Hamiltonian now has a first term $V(r)$ and a spin-orbit term $H_{SOC} = \lambda\vec{\sigma}.\vec{k}$, which is an odd function of the crystal momentum $\vec{k}$, instead of the moment, as is the case for an electron moving around a single atom. As a result, the bands split along $k$ into a spin-down band that has a lower energy in the $+k$ direction, and a spin-up band that has a lower energy in the $-k$ direction Fig. 4.3 bottom). For completion, note that SOC gives rise to many variations of this band structure, as they depend on crystal-field splitting and must obey symmetry relations. However, in all variations the bands split in k-space depending on the sign of $\vec{\sigma}$ (in the image in Fig. 4.3, the $k$ vector points along $x$ and $\vec{\sigma}$ is polarized along $y$).

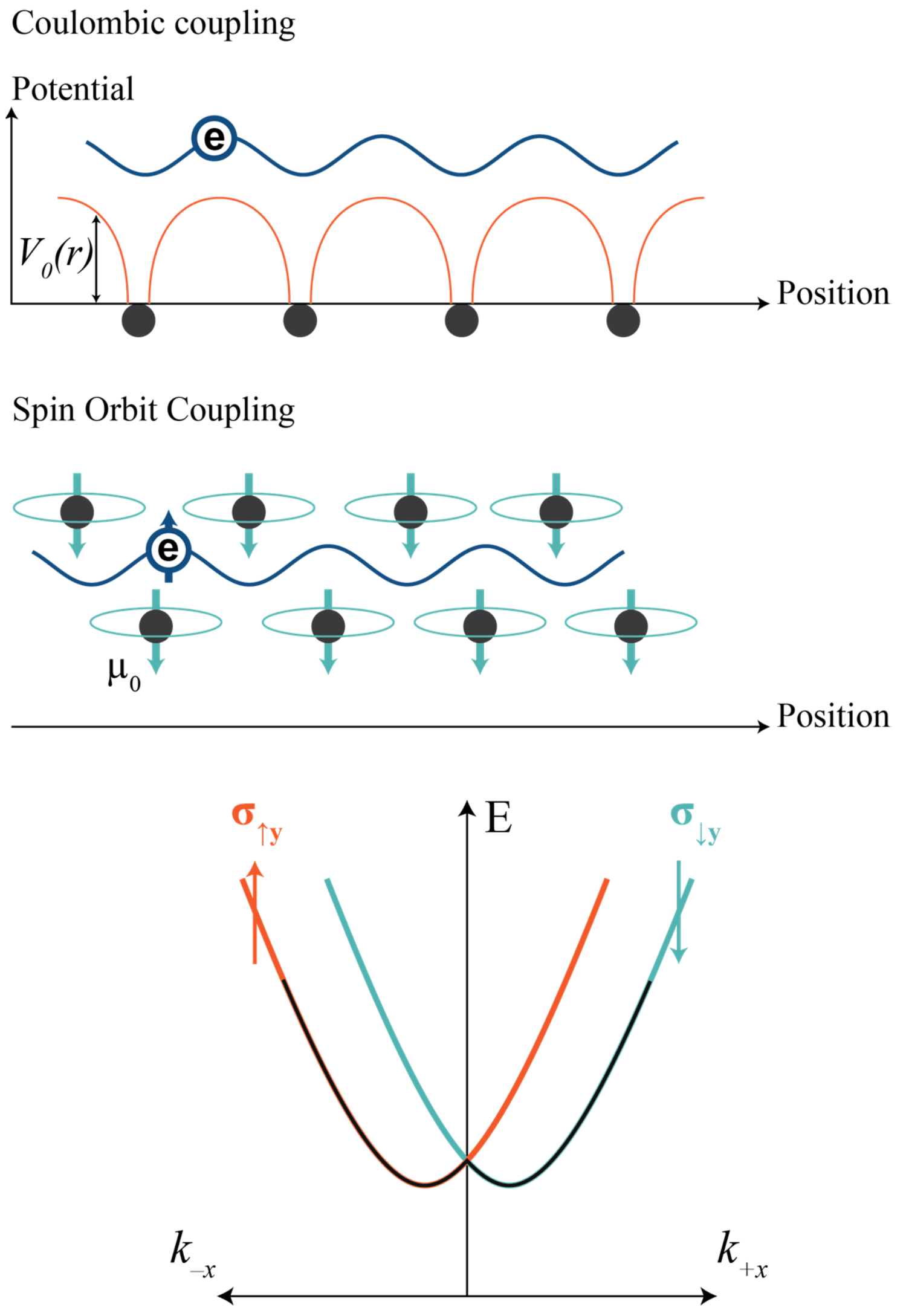

Fig.4.3 The effect of SOCs on the band structure.

The split bands, in turn, give rise to the SHE and ISHE, as shown in Fig. 4.4[49], via $\vec{\sigma}$-dependent scattering of the electrons. An external magnetic field along $y$ defines $\vec{\sigma}$. Applying an electric field $\vec{E}$ along $x$ results in a shift of the bands by a drift velocity parallel to $k_x$, and limited by electron back-scattering from $+k_x$ to $-k_x$ as in the Boltzmann transport equation. This back-scattering is accompanied by a decrease in $\sigma\downarrow$ electrons and an increase in $\sigma\uparrow$ electrons, a net spin polarization. Thus, it gives rise to a transverse spin flux $j_{Sz}$ that creates a transverse spin accumulation along the third direction $z$. Because of the analogy with the AHE, this new effect also took the Hall name to become the SHE.

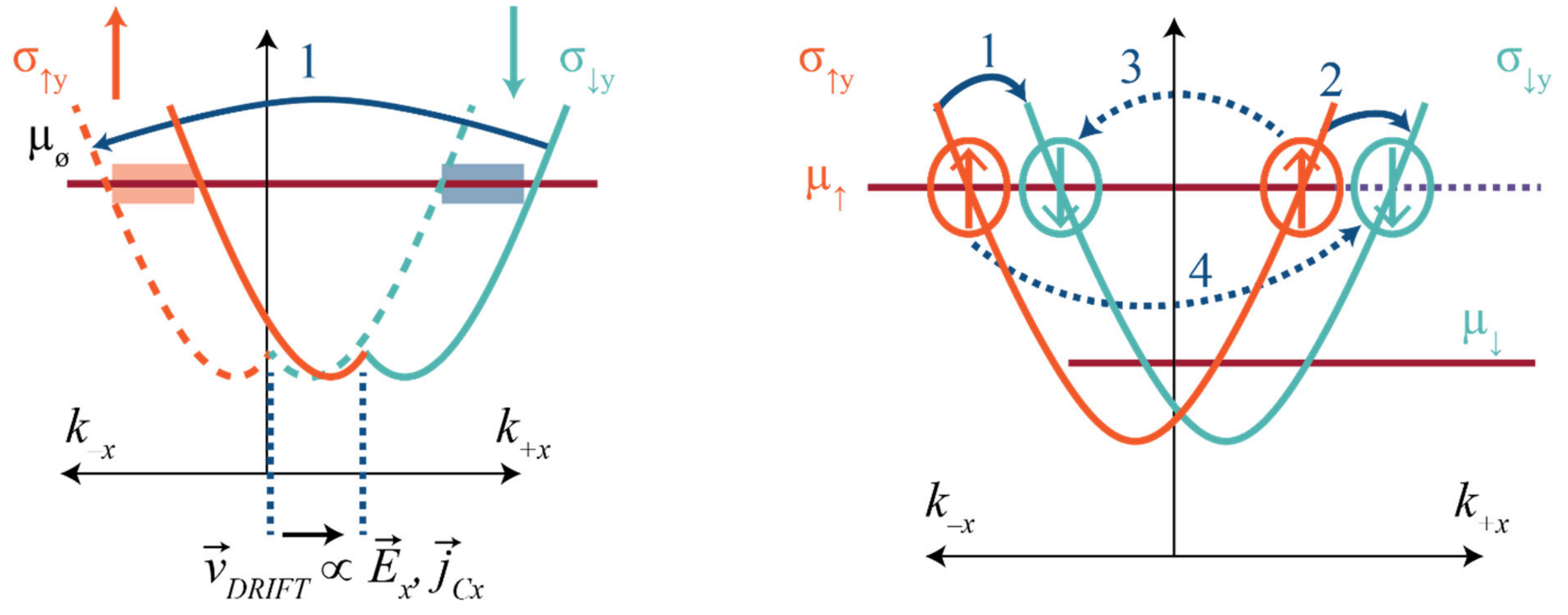

Fig. 4.4 The origin of the intrinsic SOC SHE (left) and ISHE (right)[49].

The intrinsic SHE was observed in GaAs by Kato et al.[48], who used the optical Kerr effect to detect spin polarization in GaAs (see Fig. 4.5, where the axes were relabeled vis-à-vis the original publication in order to correspond to Fig. 4.4).

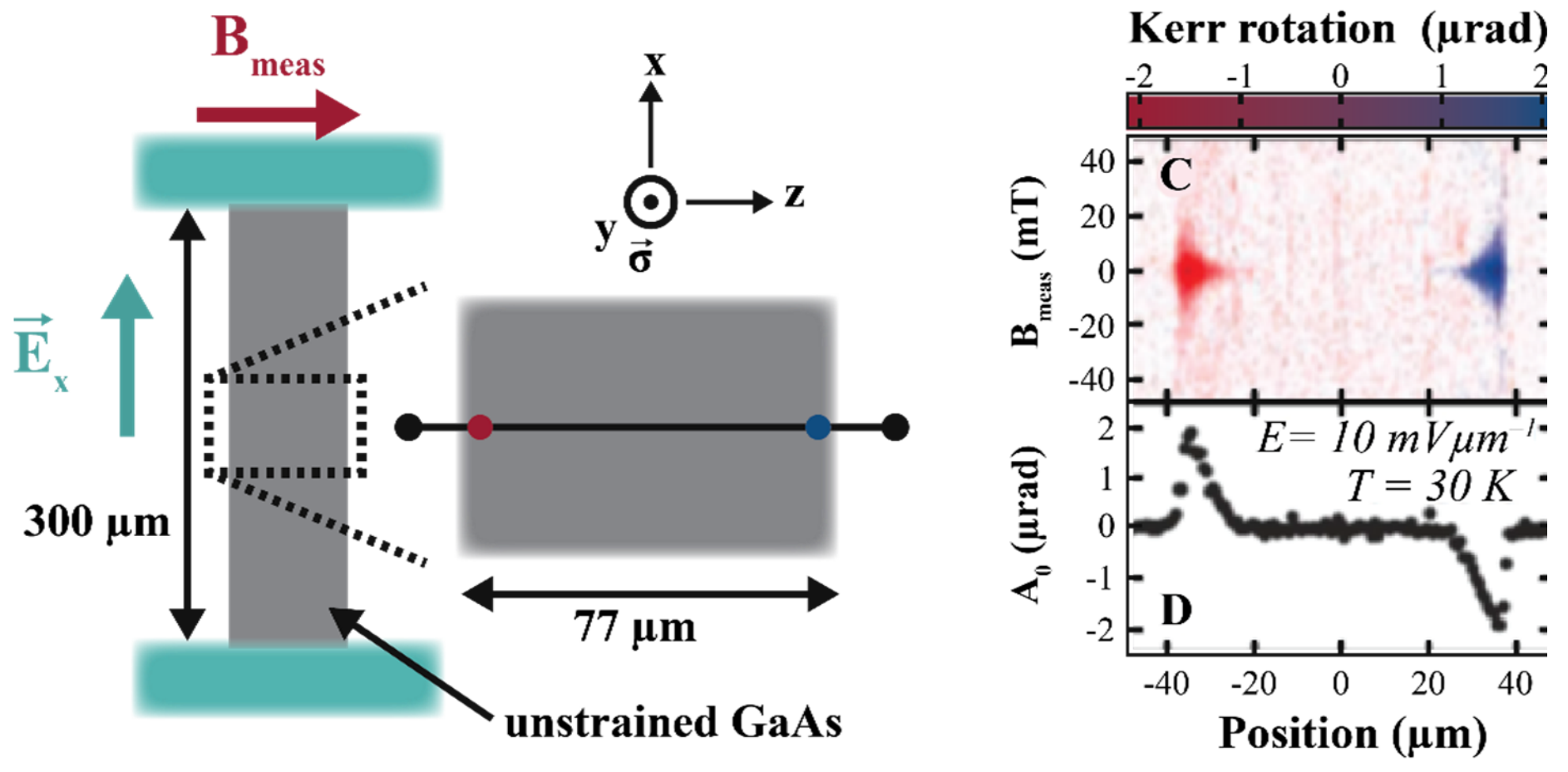

Fig. 4.5 Direct observation of the SHE in GaAs. Adapted from Ref. 48.

In this experiment, polarized light is incident onto a sample. Its reflection on spin-polarized electrons creates a small rotation in the polarization of the reflected light, the Kerr rotation, which is detected and used as a measure of spin polarization. The figure clearly shows that when current is applied in the $x$-direction of the sample and a polarizing field sets $\vec{\sigma}$ along $y$, a spin polarization appears along $z$. Since no spin current is allowed outside the sample, the spin polarization accumulates at the sample edges. The spin chemical potential $\mu_S = \mu_\uparrow - \mu_\downarrow$ (Eq. 2.3) is congruent with the curve $A_0$ in Fig. 4.4. If the same measurement had been taken with a closed spin circuit, e.g., if the sides of the sample had been coated with a spin-absorbing material, a spin flux $j_{S,z}$ would have appeared. The ratio between this spin current and the charge current that drives it gives the spin-Hall angle $\theta_{SH}$:

$$\tan(\theta_{SH}) \equiv \frac{j_{S,z}}{j_{C,x}} \quad (4.2).$$

More details about measurement in spin-open-circuit and closed circuit are given in Boona et al.[4] A table giving $\theta_{SH}$ and $L_S$ for a large variety of solids is given by Hoffmann.[50] As a general rule, strong SOCs result in a high value for $|\theta_{SH}|$, but also in a short spin lifetime and diffusion length.

The Onsager reciprocal of the SHE is the ISHE. Here, injecting electrons spin-polarized along the $y$-direction into a sample, by injecting a spin flux $j_S$ along the $z$ direction, results in the appearance of an electric field (Fig. 4.1 E). The effect is the open-circuit equivalent of the spin-galvanic effect, where instead of a transverse field, a transverse charge current $j_C$ appears. The physics underlying the intrinsic spin-galvanic effect, and by extension ISHE, is illustrated in Fig. 4.4. The injected $\vec{j}_S$ results in an unbalance between the densities of spin-up and spin-down electrons, say so that $\mu_S = \mu_\uparrow - \mu_\downarrow > 0$. Considering now the number of possible spin-flip events from the spin-up majority carrier to the spin-down minority carrier bands (labeled 1-4 in Fig. 4.4), one notices that transitions 1, 2, and 4 all tend to impel momentum in the direction $+k_x$, with only transition 3 impelling momentum along $-k_x$. Thus, a net charge current $\vec{j}_C$ will appear in the sample in the direction of $+k_x$. Again, the ratio between spin and charge current is given by the same $\theta_{SH}$ as the SHE, Eq. (4.2). In open circuit conditions, $\vec{j}_C$ will create a charge accumulation, and, thus, an electric field $\vec{E}$, the ISHE field, which is normal to both the spin-polarization direction and the direction of the injected spin flux.

The ISHE has been measured experimentally first, as far as this author is aware, by Valenzuela and Tinkham[51] and the Saitoh group[52]. It has become the most direct, all-electrical spin flux measurement, functioning essentially like a spin-ammeter. Conversely, the SHE has become an all-electrical method to inject spin currents into a material, acting in practice like a spin-current source. Both designs require depositing a thin film of a metal with strong SOC on top of a material in which one wishes to inject or measure spin currents. Thus the transmission of spin currents across interfaces has to be discussed first; it is the object of section 5.

There also are thermoelectric (Nernst) effects associated with the intrinsic SHE and ISHE via the Mott formula, Eq. (4.1). A direct equivalent of the Kato experiment, Fig. 4.4, where a heat flux replaces the current flux, thus, predicted to give a spin-Nernst effect (SNE), has been attempted by several groups, but has not been successful to date. However, the SNE has been measured successfully indirectly.[53,54] A review of spin-based Nernst effects is given by Boona et al.[4]

## 5. Spin transport across interfaces

Interfacial magnetism is a field of study in its own right.[55] Only a few aspects that pertain to thermal spin transport, as will be described in section 6, are reviewed in this paper. Spin currents can cross metal/metal interfaces just as electrical currents do. On top of that, thermally driven spin currents actually can cross interfaces between electrically insulating FMs (and sometimes even AFMs) and metals: the spin current resides in magnons on the FM side, and is converted, by conservation of spin-angular momentum, from a magnon current into a spin-polarized electron current in the metal. These two effects are treated separately.

### *5.1 Electronic spin transport across metal FM/ NM interfaces.*

In a seminal paper, Johnson and Silsbee[56] (JS) measure and explain the generation of a spin current in a NM via the application of a voltage across the junction of a single-domain FM metal and the NM. While they were not the first to study the problem of spin-current decay in the NM (see review[57]), their results inspired significant development in the field of spin-current injection. JS use the two-fluid model (Fig. 2.1) for the spin-polarized band structure of the FM. At thermodynamic equilibrium, the electrochemical potentials of the spin-up and spin-down bands in FM and NM are all aligned (Fig. 5.1[57] top). When a current is passed through the FM metal to the NM under the effect of an applied potential (Fig. 5.1 bottom), more electrons from the $\mu\uparrow$ band are injected into the NM than from the $\mu\downarrow$ band if the DOS of the former is larger at the chemical potential (see, Fig. 5.1). The formalism of Eq. (2.1) – (2.5), applied to bulk conductivities $\sigma\uparrow$ and $\sigma\downarrow$, can be applied to interfacial electrical trans-conductances $G\uparrow$ and $G\downarrow$, of the spin-up and spin-down electrons. At the interface, Eq. (2.4) becomes:

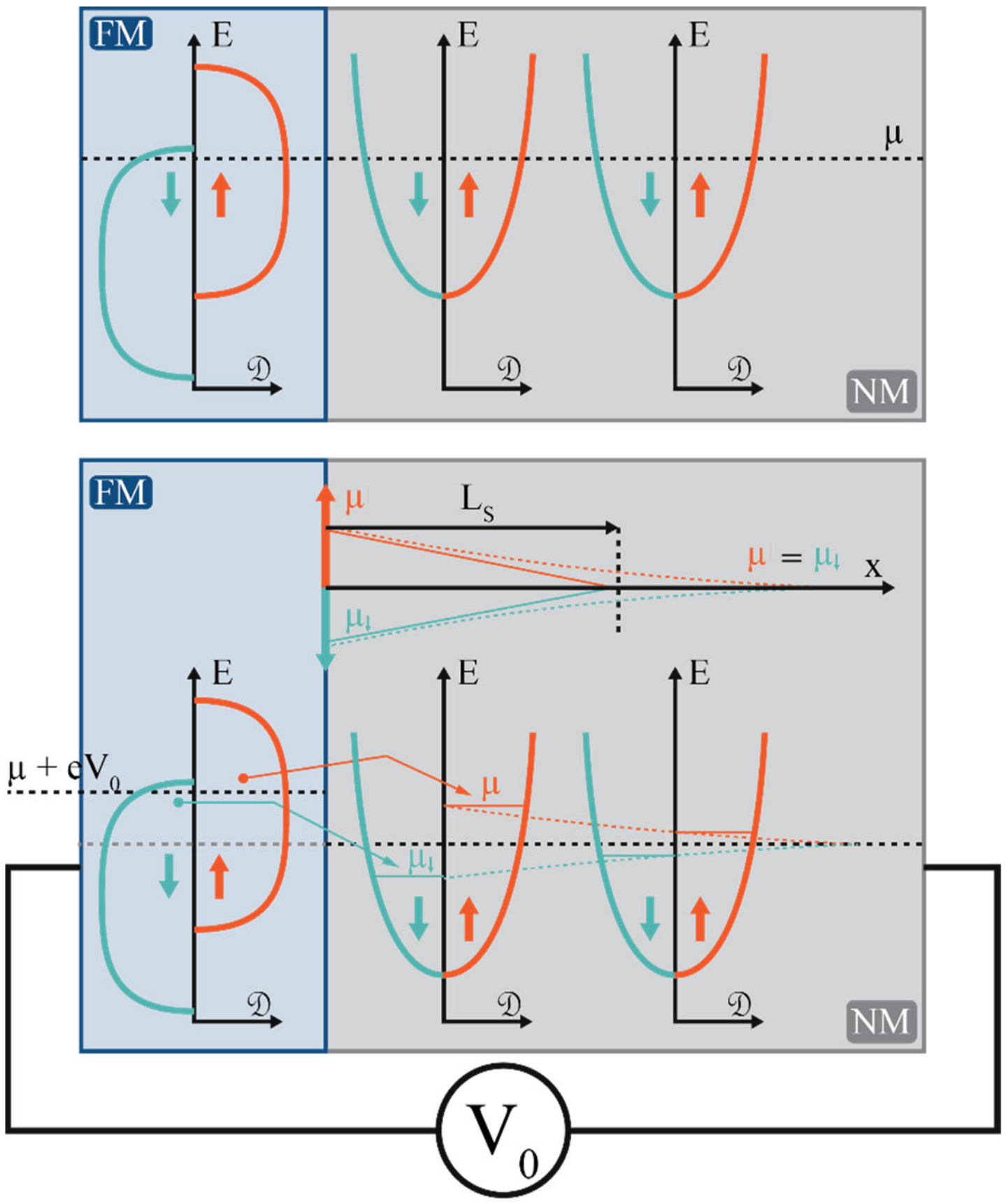

Fig. 5.1 Spin transfer torque between FM and normal metals. Adapted with permission from Ref. 57, from the Royal Society of Chemistry.

$$G = G_{\uparrow} + G_{\downarrow}$$
$$G_{\uparrow\downarrow} = G_{\uparrow} - G_{\downarrow} \quad (5.1).$$

Here, $G$ is the electrical trans-conductance, whereas $G\uparrow\downarrow$ is the *spin mixing conductance*. Thus, the spin polarization of the current in the FM is transferred into the adjacent NM: JS show that the corresponding magnetization flux $j_M$ associated with the spin current is:

$$j_M = j_C \frac{\mu_B}{e} \eta \quad (5.2),$$

where $\eta = \frac{G_{\uparrow} - G_{\downarrow}}{G_{\uparrow} + G_{\downarrow}} = \frac{G_{\uparrow\downarrow}}{G}$ is a dimensionless constant.

Once the spin flux has penetrated into the NM, it decays by spin-flip interactions following the description in Fig. 2.2 and Eq. (2.7)-(2.8). The length scale for this decay is $L_S$, which is on the order of a few nanometers in platinum and gold, but can several microns in copper and aluminum[58]. Note that, as a general rule, the materials with the strongest SOC and largest $\theta_{SH}$ also have the shortest values of $L_S$: strong SOCs promote spin-flip transitions. Such materials also act as spin-sinks: depositing a film of a high-SOC metal, like Pt, decreases the spin accumulation in the magnetic or non-magnetic material under it.

The inverse problem from Fig. 5.1, the NM/FM interface, is also described by JS. If the NM layer thickness is below $L_S$, so that there is still spin polarization in the NM, the transmission of current across the NM/FM will re-transfer this polarization to the FM, affecting its magnetization.

*5.2 Spin pumping and spin transfer torque*

Consider now the case, Fig. 5.2,[59] where the spin in the FM layer resides not in conduction electrons, but in magnons. This case applies to interfaces between NMs and both metallic and insulating FMs. The FM layer has a magnetization $\vec{\mu}$ that precesses, as shown in Fig. 5.2. This moving magnetization causes a spin-polarization of the electrons in the adjacent NM layer, in effect "pumping" a spin flux $j_S$ across the interface from the FM to the NM. The spin-polarization of the electrons in the NM arises from the conservation of spin-angular momentum across that interface. The effect can be estimated by calculating the reflection and transmission coefficients of the magnetization flux at the interfaces, in a fashion that is analogous to scattering theory valid for electron transmission across interfaces (Eq. 5.1). In this situation, the scattering is similar to *s-d* scattering in FMs, where by "*s*-electrons," we mean the conduction electrons in the NM, and by "*d*-electrons," we mean the core electrons in the unfilled *d*-shells in the FM on which the magnetization resides.

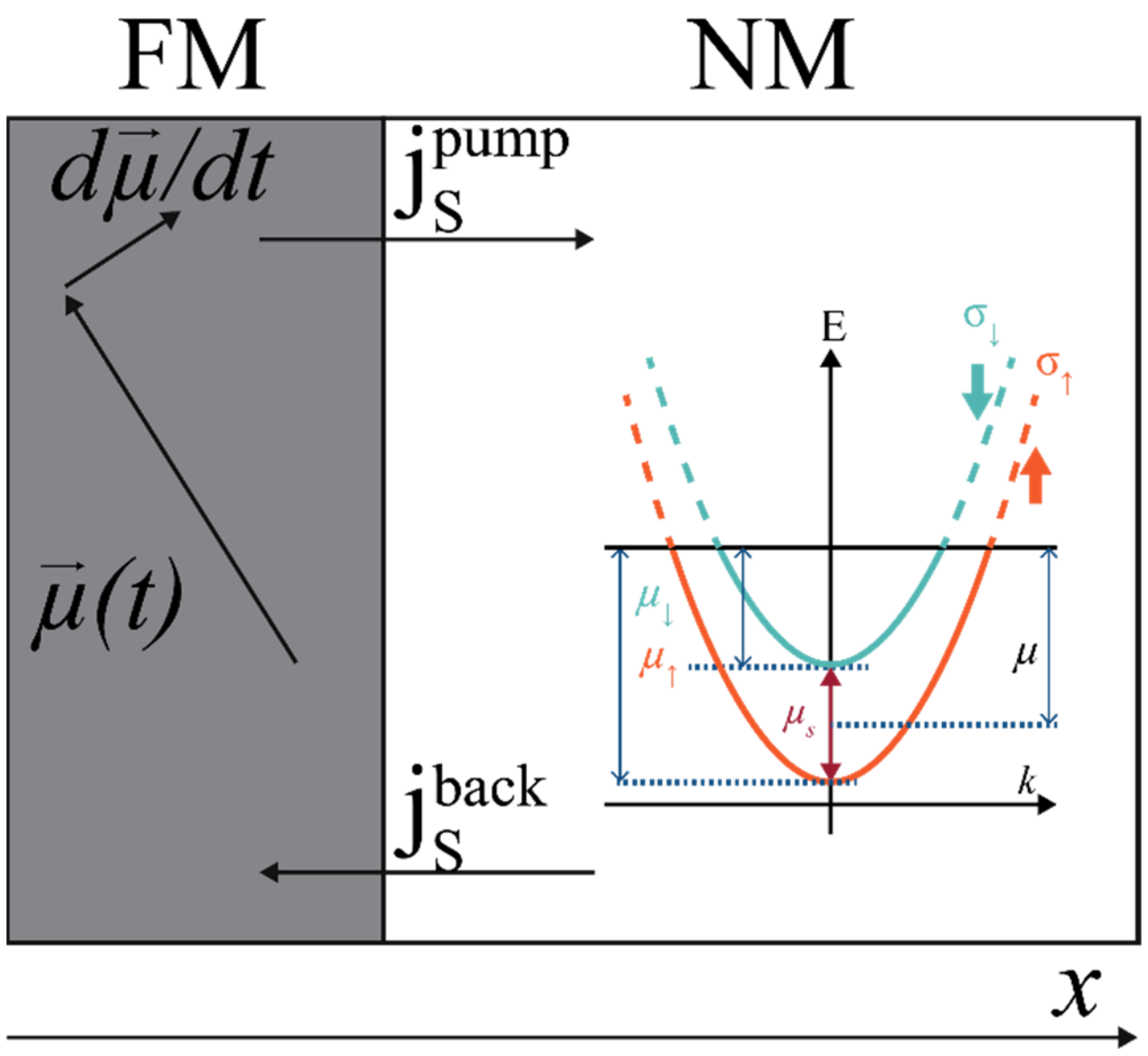

Fig. 5.2 Spin transfer torque and spin pumping between magnons in an FM and electrons in a NM layers. This applies to both metallic and non-metallic FMs. Adapted with permission from Ref. 59, copyrighted by the American Physical Society.

Once again, the Onsager reciprocal of spin pumping exists, known as *spin transfer torque* (STT).[60] Consider the case where the FM in Fig. 5.2 is an electrical insulator, so that only magnons can support a spin flux in it. A spin polarization in the NM, induced either by passing a charge current through a FM metal or by inducing an ISHE in a metal with high SOCs, will transfer spin torque to the FM, and induce a spin flux $j_S$ carried by magnons, the STT.

*5.3 Designing spin current sources and measurements*

Combining spin pumping with ISHE makes it possible to design all-electrical spin flux detectors. Conversely, combining SHE with STT makes it possible to design all-electrical spin-flux sources. Any detector or source design must keep the directions of the spin polarization (typically via an applied magnetic field or spontaneous magnetic moment), the spin current or flux, and the electric field or current orthonormal to each other.

All-electrical spin detectors can be designed to measure a spin flux moving from an insulating material into an adjacent NM film with strong SOC. For example, suppose a 7 nm ($< L_S$) thick Pt or W NM-film with large $\theta_{SH}$ is deposited on a sample. The spin-flux propagation direction must be normal to the plane of the film. Both the spin polarization (i.e., the applied magnetic field) and the ISHE voltage to be measured must be in the plane of the NM-film, but normal to each other. The spin flux will cross the sample/Pt-film or sample/W-film interface (see section 5) and spin-polarize electrons in the Pt or W. This in turn will generate an ISHE field that is measured. Pt and W have opposite signs of $\theta_{SH}$. This sign change can be used to test that any measured voltage indeed arises from a spin flux and ISHE: the polarity of the voltage signals should change sign when the film is changed from Pt to W. The film thickness has to be maintained at or below $L_S$; otherwise, the fraction of the film thickness that is above $L_S$ acts as an electrical short to the ISHE (see decay in Fig. 5.1).

This approach does not apply when the spin flux to be measured originates from a metallic FM with a higher electrical conductivity than Pt or W, because that metal will short-circuit the ISHE voltage. However, the approach can work if the spin source has a lower electrical conductivity than the detector, e.g., when it is a semiconductor with spin-polarized electrons. In principle, also, since spin currents can traverse AFM electrical insulators with long $L_S$,[61] presumably in the form of AFM magnons, it is possible to grow a thin, electrically insulating, but spin-transmitting layer between a FM and NM, and still detect an ISHE field in the NM[61].

A source that can inject a spin flux into an electrically insulating material is obtained as the Onsager reciprocal of the structure above. Again, a Pt or W NM-film is evaporated onto the material in which a spin flux is to be injected, but now one passes an electrical current through the NM film. In the presence of a magnetic field perpendicular to the current, this causes the injection of a spin current $j_S$ normal to the thickness of the strip via the SHE. Thus, the same structure can serve the purpose of a spin current source. Finally, to inject a spin current into an electrically conducting material, one can inject a charge current normal to the interface between a NM and an FM, as in Fig. 5.1.

## 6. The spin-Seebeck effect

Both the spin-Seebeck (SSE) and magnon-drag (MD) effects are advective transport processes involving two separate fluids; here, either magnons and electrons, or electrons in two separate solids, one with spin-polarized electrons and the other a NM with ISHE. The similarity between SSE and MD was first pointed out by Lucassen and Duine.[62]

The longitudinal geometry for measuring the SSE is shown in Fig. 6.1. It is very similar to the geometry one would use to measure the Nernst effect on a bulk sample; therefore, it is only applicable to FM insulators as any free electron in the FM would give rise to a Nernst voltage that would contaminate the SSE signal. The most studied FM insulator is YIG (actually a ferrimagnet in which the Fe atoms on octahedral sites have a magnetic alignment opposite to that of the Fe atoms on the tetrahedral sites, but with a net moment nonetheless). The spin current in the FM is supported by magnons. A temperature gradient is applied to the FM insulator along the $z$ direction in Fig. 6.1. Because heaters, heat sinks, and thermometers connect only with the phonon bath, the heat then is transferred from the phonons to the magnon system by phonon-magnon scattering. The characteristic length for this process[63] is of the order of 200 nm in YIG. Once the heat flux is in the magnon system, $\vec{j}_Q$ generates a spin flux $\vec{j}_S$ by Eq. (3.12) (Fig. 6.1). An external magnetic field sets the spin polarization $\vec{\sigma}//\vec{H}$ along the $y$ direction (Fig. 6.1). A NM (Pt) layer thinner than $L_S$ (typically 5 to 10 nm thick for Pt) is applied to the FM insulator. Spin pumping (section 5.2) spin-polarizes the electrons in the Pt, giving rise to an ISHE field long the *x* direction that is picked up as a voltage, as shown. The total structure transforms a temperature difference into a voltage, the SSE.

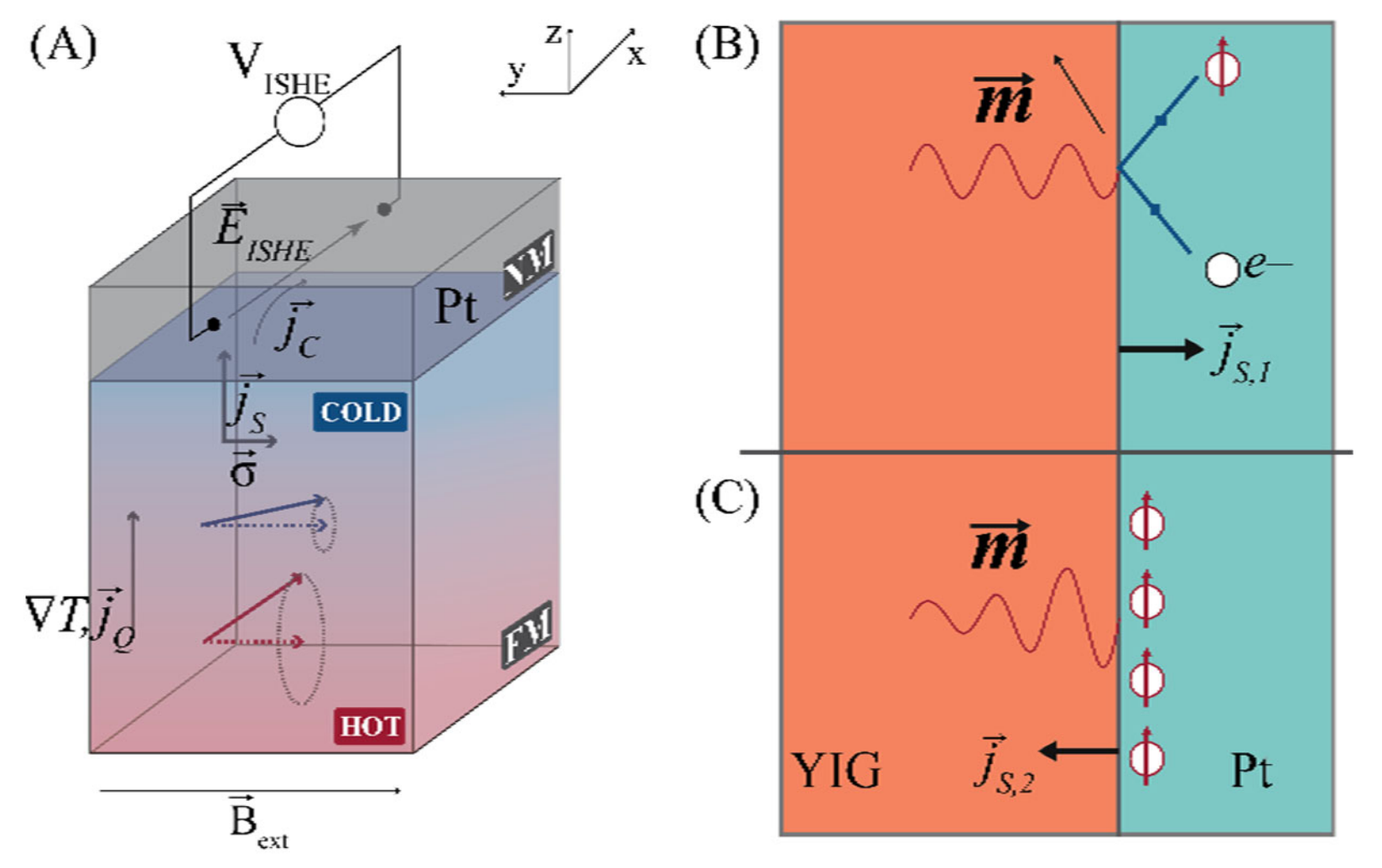

Fig. 6.1 Spin Seebeck (A) and (B) and Spin-Peltier (C) effects.

The SSE was measured first[3] on a Pt-metallic FM (Permalloy) bilayer in a different geometry, which was labelled the transverse SSE[4]. It then was measured in a FM semiconductor GaMnAs[9] and an insulating FM[64]. The largest effect was measured on Landau levels of a non-magnetic semiconductor (InSb).[65] At a field sufficiently high to confine all electrons on the last Landau level, these electrons are fully spin-polarized, and the SSE reaches 8 mV/K, a value that exceeds all thermoelectric effects on the bulk of the InSb by an order of magnitude. The transverse geometry allows for the use of FM conductors, but its results are easily contaminated by the effect of heat losses and the method requires a rigorously adiabatic mount. Because this is not widely available in laboratories that specialize in magnetic measurements, the transverse geometry is now abandoned in favor of the geometry in Fig. 6.1 developed in 2010.[64] Non-local measurements of spin transport in insulating FMs, driven by either electrical injection (5.3) or SSE, were performed.[66,67,68,69] Cornelissen's PhD[70] provides an excellent review. Optically induced, non-local thermal spin transport has been reported,[71] and again, a drift-diffusion model explains the data quantitatively[72]. SSE measurements also were used to measure the transmission of magnons though AFM layers deposited on FMs.[61] Both AFMs and paramagnetic solids provide a SSE signal, when the spin-polarization is provided by an appropriate external magnetic field.[73,74] The $zT$ of SSE measurements can be calculated, but is impractical (~$10^{-3}$).

The Onsager reciprocal of the SSE, the spin-Peltier effect (SPE), also was reported[75] (Fig. 6.1 C). Passing a current through the Pt cools or heats the magnons in the FM. The reciprocity between SSE and SPE[76] is the product of three separate reciprocity relations, as illustrated by comparing Fig. 6.1 B and C: (a) between SHE and ISHE (section 4.2), (b) between spin pumping and STT (section 5.2), and (3) in the spin/heat flux Onsager relations Eq. (3.14). Where the ISHE generates the voltage in the SSE measurements, the SHE generates spin polarization in the Pt at the Pt/FM interface in the SPE measurements. Where in SSE, magnons in the FM spin-polarize electrons in the Pt by spin pumping, in SPE the spin accumulation in the Pt transfers spin-angular momentum into the FM magnon system by STT. Finally, where the temperature gradient drives the spin flux in the magnon system during SSE, the magnonic Peltier coefficient Eq. (3.14) drives a temperature gradient in the SPE experiment.

## 7. Magnon Drag

Whereas the SSE generally is a very small effect (< 1 μV/K) at room temperature and has almost no potential applications in thermoelectric technology, the same does not hold for MD. The MD thermopower is often an order of magnitude higher than the regular diffusion thermopower in metals, and also dominates it in magnetic semiconductors. It is the only example where a spin-based effect is much larger than a charge-based effect. MD is also quite useful in the quest for high $zT$ materials.[77] Fig. 7.1 illustrates the similarities and differences between SSE (top) and MD (bottom) effects. The spin-flux generation mechanism is common to both: a temperature gradient, initially imposed on the phonon system, is transferred to the magnons system, where the magnon heat current also generates a spin current. The first difference is that in the SSE, the FM must be electrically insulating, whereas in MD, it must be an electrical conductor. The second difference is one of material quantity: in the SSE, the spin current goes though an interface; therefore, the structure must have a thin Pt (or other metal with strong SOC) film, whereas the MD effect is a bulk effect using the conduction electrons of the FM material itself. Whereas in SSE, the FM spin current transfers spin-angular momentum through an interface in the spin-pumping mechanism, in MD, the spin current transfers linear momentum to the electrons, thereby increasing the longitudinal electric field by a quantity $\vec{E}_{MD}$. The underlying physics is common: both spin pumping and linear-momentum transfer are caused by *s-d* scattering, which is very intense. *S-d* scattering it is the reason why the mobility in FM and paramagnetic conductors is much lower than in non-magnetic conductors. It is also quite independent of temperature, and persists to very high temperatures. The MD thermopower is then the ratio between the two collinear gradients:

$$\alpha_{md} \equiv \frac{\vec{E}_{MD}}{\nabla T} \qquad (7.1).$$

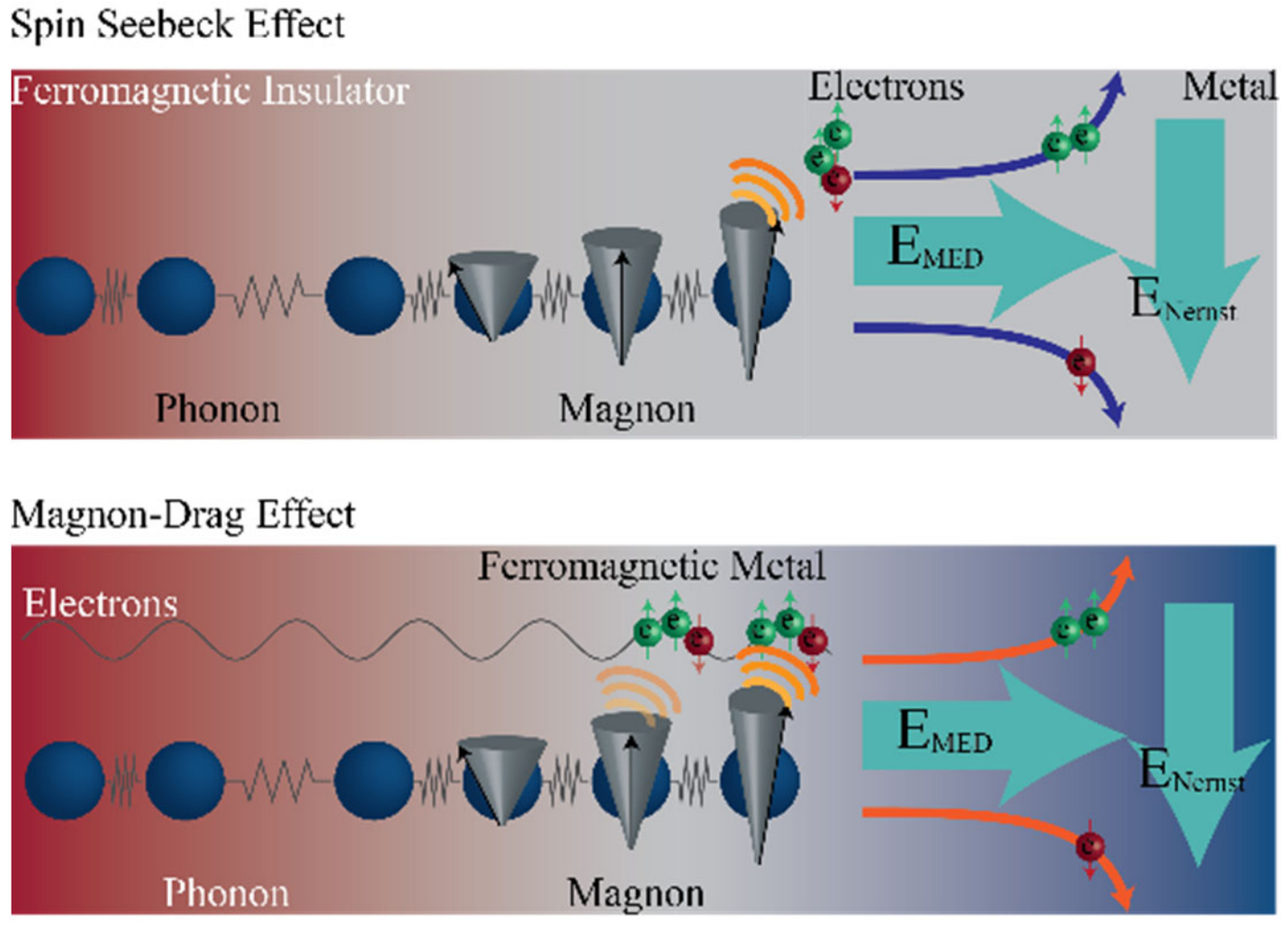

Fig. 7.1 Spin Seebeck versus magnon-drag effects

For completion, we mention that, in principle, magnons also can transfer spin-angular momentum to conduction electrons and add to the spin polarization of the free electrons in the FM, thereby adding to the electric field $\vec{E}_{Nernst}$ due to the AHE and the ISHE; this possibility, however, has not been evinced experimentally to date.

Historically, MD was suspected to be the source of the very large thermopower of iron[1] and the AFM MnTe[2]. Early theories for MD were inspired by those from phonon drag, but required excruciating calculations of scattering times, which in practice are never really accurate. Based on the modern concepts of spin transport outlined above, two practical MD theories have emerged recently.[6] One is the hydrodynamic theory based on a magnon-electron two-fluid model. The other is a spin-dynamic theory that allows for subtler predictions, but is more difficult to implement in real materials. For metals, the hydrodynamic theory is predictive and allows for the design of high-thermopower alloys.[8]

*7.1 The hydrodynamic theory of magnon drag.*

This theory considers the magnetic conductor as comprising[78] an electron and a magnon fluid. The electrons have a momentum density $\vec{p}_e = n_e m \vec{v}_e$ in terms of their number density $n_e$, mass $m$, and drift velocity $\vec{v}_e$. The magnons have momentum density $\vec{p}_m = n_m M \vec{v}_m$ expressed as a function of mass $M = \frac{\hbar^2}{2D}$ ($D$ is the magnetic exchange stiffness, Eq. 3.7), velocity $\vec{v}_m$, and density $n_m$. The equations of motion for the two coupled fluids, assuming quadratic dispersions and thus Galilean invariance, are:

$$\begin{aligned} \frac{d\vec{v}_e}{dt} &= \frac{e}{m}\left(\vec{E} - \alpha_d \vec{\nabla} T\right) - \frac{\vec{v}_e}{\tau_e} - \frac{\vec{v}_e - \vec{v}_m}{\tau_{me}} \\ \frac{d\vec{v}_m}{dt} &= -\frac{\alpha_m}{M}\vec{\nabla} T - \frac{\vec{v}_m}{\tau_m} - \frac{\vec{v}_m - \vec{v}_e}{\tau_{em}} \end{aligned} \qquad (7.2),$$

where $\vec{E}$ is the electric field and $\tau_e$ and $\tau_m$ are transport scattering mean free times for the electrons and magnons, respectively. The ordinary electronic diffusion thermopower in a metal is $\alpha_d$, given by:

$$\alpha_d = \left(\pi^2/3\right)\left(k_B T/e\right)\left(k_B T/\mu\right) \qquad (7.3).$$

The magnonic thermopower $\alpha_m$ is derived in (3.18). The time scales $\tau_{me}$ (the scattering time of electrons on magnons) and $\tau_{em}$ (the scattering time of magnons on electrons) parametrize the magnon-electron collision rate. Per the conservation of linear momentum, $\frac{n_e m}{\tau_{me}} = \frac{n_m M}{\tau_{em}}$.

Under steady-state conditions and for zero electric current ($v_e = 0$), the above equations are solved to determine the electric field required to counteract the thermal gradient. The total electron thermopower, including both the diffusive and magnon-drag contributions, becomes:

$$\alpha \equiv \frac{|\vec{E}|}{|\vec{\nabla} T|} = \alpha_d + \frac{2}{3}\frac{C_m}{n_e e}\frac{1}{1+\frac{\tau_{em}}{\tau_m}} \tag{7.4}$$

and the magnon-drag contribution, isolated, is:[6]

$$\alpha_{md} = \frac{2}{3}\frac{C_m}{n_e e}\frac{1}{1+\frac{\tau_{em}}{\tau_m}} \tag{7.5}$$

In the hydrodynamic theory, the frictional forces between magnon and electrons push the electrons, alongside the magnons, toward the cold side of the sample, giving a thermopower that has the sign of the effective mass of the main charge carrier times the charge of the electron, i.e., $\alpha_{md} < 0$ when $m > 0$.

The factor $\left(1+ {}^{\tau_{em}}\!/_{\tau_m}\right)^{-1}$ in Eq. 7.5 contains the ratio between $\tau_{em}$, the magnon scattering time for collisions with electrons, and $\tau_m$, the total magnon scattering time, which includes collisions with all scatterers, electrons, defects, phonons, or other magnons. The scattering ratio $\tau_{em}^{-1} / \tau_m^{-1}$ represents the efficiency with which magnons transfer their momentum to electrons. The scattering ratio is different depending whether the solid with MD is a metal, where it is about unity, or a semiconductor, where it can be of order $10^{-2}$. It also varies with temperature.

We consider the total magnon scattering frequency $\tau_m^{-1}$ first. Its temperature dependence is different below and above a threshold temperature $T^*$. At higher temperature ($T > T^*$), magnon scattering is likely to be taken over by Gilbert damping[79] in semiconductors, where there are less than $10^{-3}$ free electrons per atom. It is then parametrized by the dimensionless constant $\gamma$ as $\tau_m^{-1} \propto \gamma T$. The crossover takes place at $T^* \sim \frac{T_c}{s^{2/3}}(\gamma l)^2$. Here, $s$ is the saturation spin density, about 1 per atom, so that $s \sim a^{-3}$. Using $\gamma \sim 10^{-2}, a \sim 1nm$ and scattering length $\ell_m \sim 1$ μm, we obtain a range of $T^* \sim (10^{-1}$ to $10^{-3})T_c$ (for iron, $T^* \sim 100$ K). At lower temperature ($T < T^*$) and in semiconductors, we assume an energy-independent disorder-dominated magnon mean-free path $\ell_m$, and then we expect that $\tau_m$ scales with temperature as $\tau_m^{-1} \propto \sqrt{T}$ because the FM-magnon DOS scales as $\mathfrak{D} \propto \sqrt{\hbar\omega}$. In metals, conductors with about one free electron per atom, magnon scattering is dominated by electrons $\tau_m^{-1} \sim \tau_{em}^{-1}$ at low temperature.

The electron-magnon scattering frequency $\tau_{em}^{-1}$ is expected to scale with temperature as $\tau_{em}^{-1} \propto T^2$. This results from the combination of momentum and energy conservation constraints for electron-magnon scattering, which give a factor of $\sqrt{T}$, and the reduced phase space for occupied magnon states, which gives a factor of $T^{3/2}$.

In all cases, the factor $\left(1+\tau_{em}/\tau_m\right)^{-1}$ should tend to unity with temperature as temperature approaches zero. In semiconductors, in the Gilbert-damping dominated limit, the attenuation of the MD thermopower is expected to have a linear dependence on $T$. Conversely, in the regime where $\tau_m$ is dominated by magnon-phonon scattering, $\tau_m^{-1}$ vanishes faster than $\tau_{em}^{-1}$ due to the rapidly shrinking phase space, and the factor $\left(1+\tau_{em}/\tau_m\right)^{-1}$ approaches unity. In metals, the factor $\left(1+\tau_{em}/\tau_m\right)^{-1}$ also tends to unity (it becomes ½) as $\tau_m^{-1} \sim \tau_{em}^{-1}$. Thus, except for the case of semiconductors at high temperature, $\alpha_{md}$ should follow the $T^1$ to $T^{1.5}$ law of $C_m$ in FMs, or $T^3$ in AFMs.

*7.2 Magnon drag due to internal spin pumping*

The spin-dynamic MD theory considers the motion of the electrons over a magnetically textured landscape, where the texture comes from the presence of magnons. The electron spins then track this texture in a way similar to the calculations of spin-orbit interactions in Fig. 4.3, but because this texture is magnon-induced and dynamic, it gives the electrons a dynamic magnetization. The MD thermopower is calculated from the electric current pumped by the magnetization associated with a magnon heat flux $\vec{j}_Q$ as calculated by Lucassen et al.:[62]

$$\vec{j}_C = \sigma(\vec{E} + \beta p \frac{\hbar}{2e} \frac{\vec{j}_Q}{sD}) \quad (7.6).$$

Here, $\beta$ is a dimensionless coefficient (typically around 0:1 to 0:001) quantifying the lack of spin conservation in the interaction between spin current and magnetization dynamics, and $p$ is the spin polarization of the electric current (typically of order 1). $D$ is the stiffness, Eq. (3.7). The thermopower is then:

$$\alpha_{md}' = \beta p \frac{\hbar}{2e} \frac{\kappa_m}{sD} \quad (7.7)$$

Eq. (7.5) is rooted in purely nonrelativistic Galilean momentum transfer between magnons and electrons, while Eq. (7.7) is based on spin-orbit interactions, an effect based on the relativistic Hamiltonian $H_{SOC} = \lambda \vec{\sigma}.\vec{k}$. Remarkably, the estimates in Eqs. (7.5) and (7.7) coincide if we set the scattering factor in (7.5) to unity and set $s/n_e$=1 and $\beta p$=1, a reasonable assumption for iron, cobalt, or nickel[6]. In particular, the sign of $\alpha_{md}$ and $\alpha'_{md}$ is the same.

The internal spin-pumping theory requires fewer assumptions than the hydrodynamic theory. In particular, if $\beta < 1$, expressing a lack of spin conservation in magnon transport, the transport becomes non-hydrodynamic in nature. There also are transport regimes in which the hydrodynamic and spin-pumping MD thermopowers give very different predictions, in particular about the sign of the MD thermopower. For example, Eq. (7.7) considers only the frictional forces between electrons and magnons as contributing to the advective transport process. A subsequent theory paper[80] includes the effect on the thermopower of the magnon dissipation mechanisms. This adds a second force to magnon-on-electron interactions related to the solid angle subtended by the magnon precession that pushes the magnons towards the hot side. Including this force, the MD thermopower of (6.7) becomes[80]:

$$\alpha'_{md} = \left(\beta - 3\alpha_G\right) p \frac{\hbar}{2e} \frac{\kappa_m}{sD} \qquad (7.8).$$

The sign of $\alpha'_{md}$ now can change when the effect of magnon decay (expressed by the Gilbert damping term $\alpha_G$) dominates over the frictional forces (the $\beta$ term). The ratio $\beta/\alpha_G$ equals[81] the ratio $s_t/s_i$ between the total amount of spin-angular momentum $s_t$ present in the system and the amount $s_i$ residing on delocalized electrons. This is proportional to the ratio of the net spin-polarization residing on conduction electrons and the total spin polarization, including the fraction that resides on the unfilled d- or f-levels.

*7.3 Magnon drag in metals*

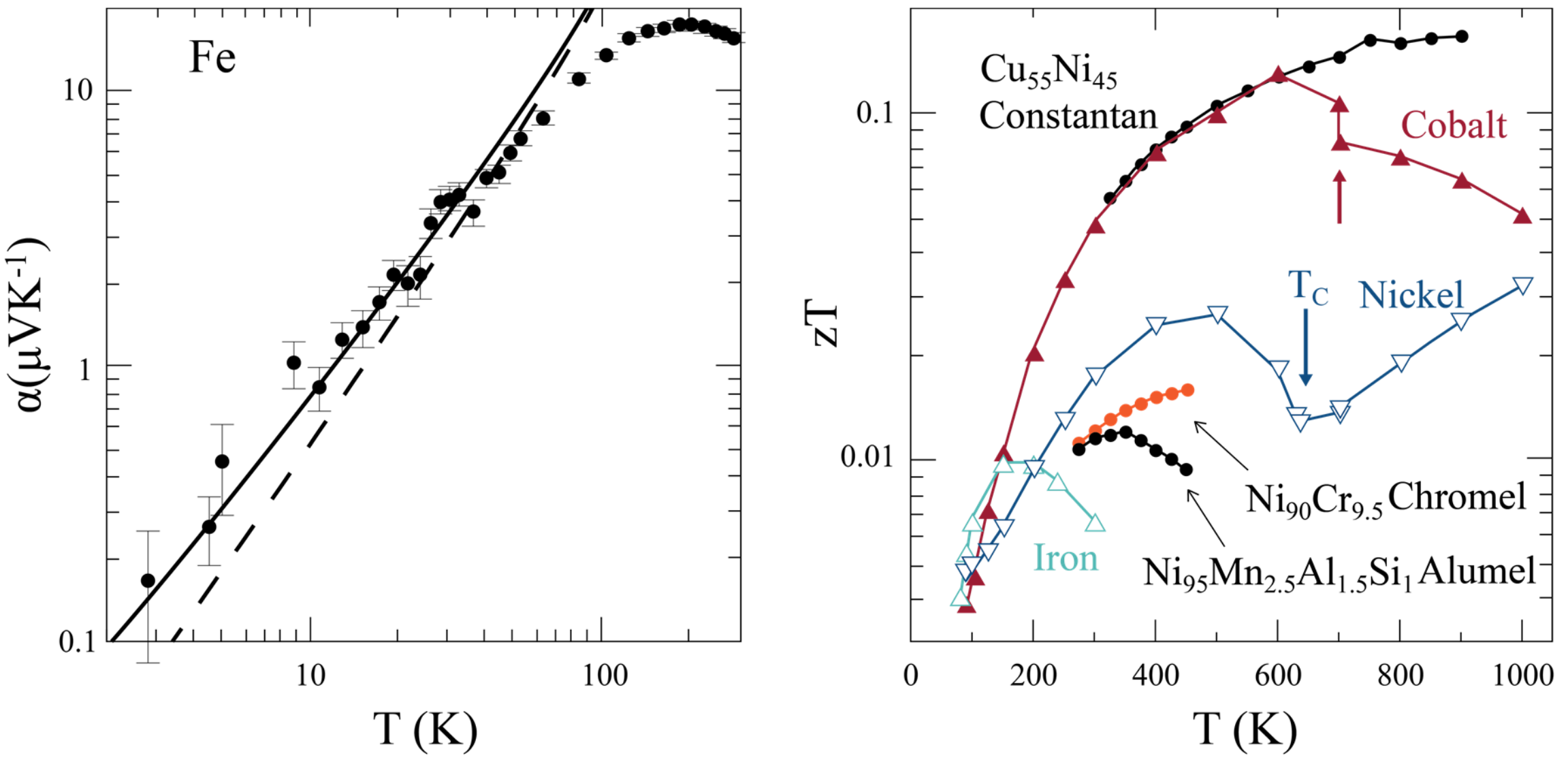

Fig. 7.2 Magnon-drag in metals: Left: Magnon-drag thermopower of iron (dashed line)[6]. Right: Some of the highest thermoelectric figures of merit achieved in metals.[7] The magnon-drag contribution to the thermopower of iron, cobalt and nickel makes these three elements and some of their alloys the metals with the highest *zT*. The arrows indicate phase transitions.

Eq. (7.5) and (7.7) calculate the thermopower of the elemental metals Fe, Co and Ni very well and without having to use any adjustable parameter:[6] Fig. 7.3 shows the thermopower of Fe, which is positive. The dashed line is the MD contribution (Eq. 7.5) calculated from the known magnon dispersion and calculated specific heat, as well as the known charge carrier concentration, and setting the factor $\left(1+\tau_{em}/\tau_m\right)^{-1}$ to unity, as justified above. Adding the diffusion term (Eq. 7.4) gives the full line, which reproduces the data up to $T^*$. Fig. 7.2 also gives the $zT$ of the metals that are known to have the highest $zT$. Except for the case of Cu-Ni alloys, the high $zT$ of all these metals and alloys are attributable to MD because, via the Mott formula, the thermopower of metals is of the order of $\alpha_d \sim \left(k_B/e\right)\left(k_BT/\mu\right) \sim 1\mu\text{V/K}$ so that for metals without MD, $zT \sim \alpha_d^2/L_0 \sim 10^{-5}\text{to}10^{-4}$. Recently, the theory also has been proven to be predictive.[82] DFT is very good at predicting electron DOS and charge-carrier density, and magnon dispersions can be calculated. Therefore, it is possible to design metal alloys to have specific values of $\alpha_{md}$, as long as $\tau_m^{-1} \sim \tau_{em}^{-1}$, i.e., when magnon scattering by free electrons dominates. Thus, the $\alpha_{md}$ of bcc Fe-Co alloys was predicted and verified experimentally.[82]

While the $zT$ of metals, shown in Fig. 7.2, is limited still to $zT \leq 0.16$, the power factor of metals is much higher than that of thermoelectric semiconductors, with thermopowers reaching 60 μV/K.[7] In fact, the thermopower of cobalt reaches 160 μW/cm $K^2$ between 300 and 400 K. There is one class of applications where the combination of a high power factor and a high thermal conductivity is very advantageous: cooling devices that operate above ambient temperature, such as electronic devices and batteries.[83] In refrigeration applications, heat backflow from the ambient temperature to the device that must be cooled is an additional parasitic load. However, in cooling applications, heat that flows spontaneously between a load that operates above ambient to ambient is beneficial to the operation of the device. Therefore, classical $Bi_2Te_3$-based Peltier coolers are not as good at cooling devices such as CPUs as a Peltier cooler made from thermoelectric metals would be. A metal Peltier cooler already provides good cooling performance passively, and adding an active Peltier heat flux to that enhances the effect manifold.[83] MD-metals, such as elemental cobalt, are suited ideally for such application.

Finally, it is noteworthy that MD, unlike phonon drag, is a high-temperature effect. In phonon drag, phonon momentum is dissipated easily into interactions with other phonons and defects before it is transferred to electrons. This is because electron-phonon coupling is relatively weak, while phonons are subjected strongly to defect and Umklapp scattering at temperatures as low as 1/10th of their Debye temperature. This limits the operating temperature of phonon drag to typically below 77 K, since most materials have a Debye temperature between 100 and 300 K. In contrast, the magnon-electron interactions that drive the MD thermopower are mediated by *s-d* scattering, which is dominant in magnetic metals. As an illustration of the relative strength of phonon-electron and magnon-electron interactions, we point out that the electron mobility in magnetic materials is typically an order of magnitude smaller than in similar non-magnetic materials. That is because the mobility in magnetic materials is limited by *s-d* scattering, whereas in NMs, phonons are the dominant electron scatterers. As a result, MD remains dominant to the ordering temperature, typically above 1000 K for most FM metals, although the range of applicability of Eq. 7.5 is limited to $T^*$, about one tenth of the ordering temperature. The more sophisticated Eq. 7.8 must be used above $T^*$. As will be shown next, MD boosts the thermopower of magnetic systems even above the ordering temperature.

<u>*7.4 Magnon drag and paramagnon drag in semiconductors*</u>

Good thermoelectrics are degenerately doped semiconductors with typically $10^{-3}$ to $10^{-4}$ free charge carriers per atom. MD applies to that situation as well, and the factor $1/n_e$ in Eq. (7.5) should result in a high $\alpha_{md}$, if it were not for two unfortunate facts. First, many mechanisms other than free electrons scatter magnons in semiconductors so that $\tau_m < \tau_{em}$. Besides decreasing $\alpha_{md}$ and thus $zT$, this also means that Eq. (7.5) loses its predictive quality because scattering times are notoriously difficult to calculate. Second, there are no FM semiconductors known with ordering temperatures $T_C$>80 K. There are AFM semiconductors with $T_N$ above room temperature, but we know of none where $T_N$ ~1000 K. Thus, the discovery of *paramagnon drag* (PMD) was critical.[8]

Fig. 7.3 shows the thermopower of Li-doped MnTe samples. MnTe, a hexagonal crystal, is an AFM with magnetic sublattice orientations that are FM in the hexagonal planes, but AFM-like between planes. The ordering temperature is $T_N$ = 305 K. At $T < T_N$, the thermopower follows a functional $\alpha(T) = \alpha_1 T + \alpha_2 T^3$ where $\alpha_1 T$ is the diffusion thermopower and $\alpha_2 T^3$ can be fitted to Eq. (6.5), which is also valid for AFMs, with a single parameter $\tau_{em}/\tau_m \sim 100$ for all temperatures and charge carrier concentrations.

What is surprising is that the thermopower in the paramagnetic (PM) regime at $T > T_N$ does not decrease back to the $\alpha_1 T^1$ law valid for the diffusion thermopower, which should extrapolate to zero in the limit for $T \rightarrow 0$.[8] Instead, at $T > T_N$, the thermopower follows a law $\alpha(T) = \alpha_1 T + \alpha_{PMD}$ where $\alpha_{PMD}$ is $T$-independent, but increases with decreasing $n_e$. Inelastic neutron scattering data show the existence of a spin structure in PM-MnTe at the energy where the AFM magnon band was in AFM-MnTe, with a temperature-independent spin-spin correlation length $\xi \sim 2.3 \pm 0.2$ nm, and a spin lifetime of $\tau_{L,PM} \sim 27 \pm 1$ fs. This spin structure is known as a *paramagnon.* Because $\xi$ is much larger than the electron de Broglie wavelength, $1/k_F \sim 0.6$ nm, and because the spin lifetime is much longer than the electron scattering time calculated from mobility $\tau_e \sim 3 \pm 2$ fs, the authors suggest that, to the electrons, this paramagnon looks like a fully developed magnon that can give an MD thermopower $\alpha_{PMD}$. Surprisingly, the only $T$-dependence of $\alpha$ comes from $\alpha_d$, while $\alpha_{PMD}$ appears to not be temperature dependent at least to $T = 3\ T_N$, except in the lowest-doped sample.

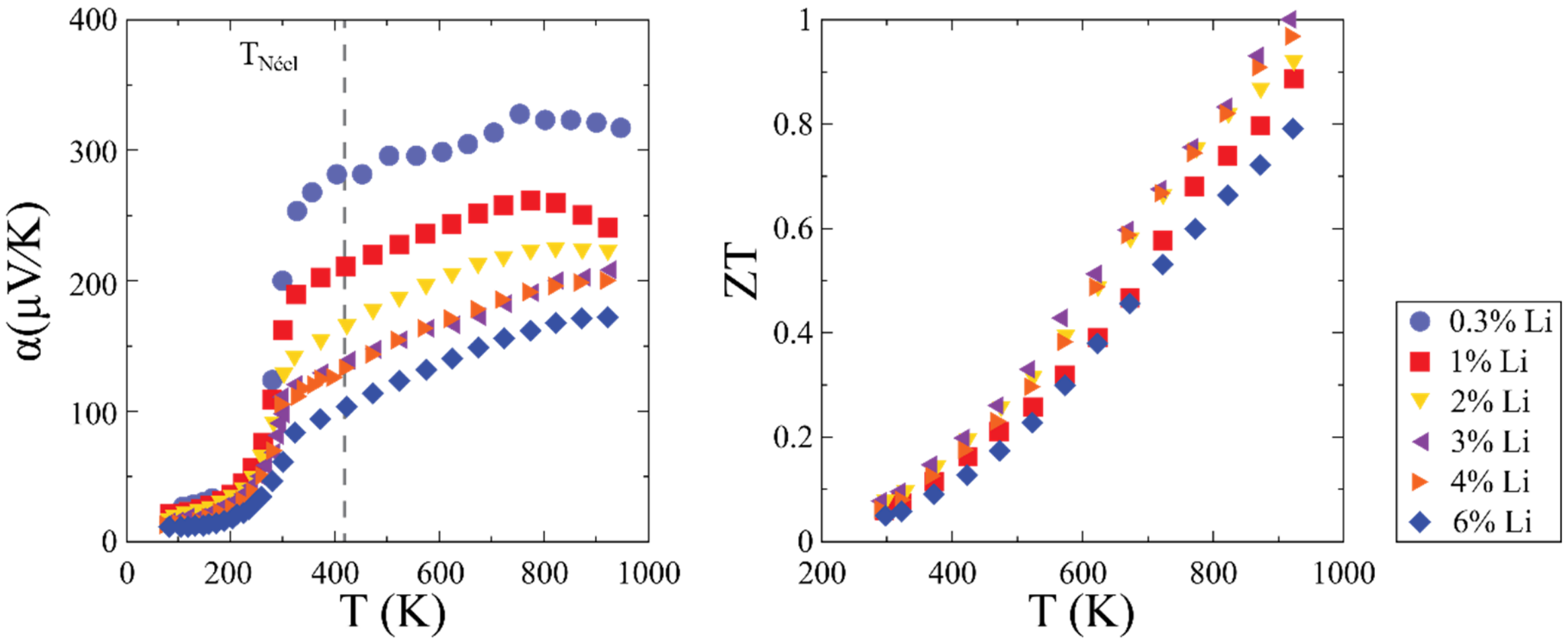

Fig. 7.3 Magnon drag in semiconducting AFM MnTe doped with Li.[8] Left: thermopower below the Neel temperature ($T_N$ = 305K) follows a mostly $T^3$ law but maintains a very high value at $T>T_N$, due to paramagnon drag. Right: this allows for a very high $zT$ to be reached in these simple binary semiconductors in their paramagnetic regime.

Paramagnons typically consist of a set of a few AFM-coupled spins, (e.g., in a 1D picture ↑↓↑), extending spatially over $\xi$. The paramagnon-neutron scattering signature has no $T$-dependence at all to 450 K (no data are available above that). A simple physical picture for this lack of temperature dependence is based on the fact that to destroy the spin-spin correlation in a short-range ↑↓↑-spin structure requires flipping a single spin 180°. This is because, unlike the case of long-range magnons, there simply are not an unlimited number of aligned spins available over which to spread the decrease of magnetization like in ordered solids (see Fig. 3.1). The energy it takes to flip the middle spin in such a structure from ↑↓↑ to ↑↑↑ is much higher than the exchange coupling energy $J$ that binds the long-range collective spin waves at $T < T_N$, because to flip a single spin, one must count the interactions of that spin with all its nearest (or next-nearest) neighbors. Because each Mn in MnTe has 4 nearest neighbors, flipping a single spin 180° requires 4 times more energy, and thus $\xi$ is not going to decrease before the temperatures reaches 3 to 4 $T_N$.

The excess thermopower ascribed to PMD makes it possible to reach $zT > 1$ in this simple binary semiconductor without any additional optimization of grain size or nanostructuring to reduce thermal conductivity. Note that PMD is a higher-temperature effect yet than MD. It also extends the number of semiconductors that are candidate high-$zT$ materials. In the past, adding magnetic ions to a semiconductor was recognized to be detrimental to the charge carrier mobility, because magnetic *s-d* scattering is extremely effective and reduces mobility. The new knowledge that this very same interaction can result in the appearance of PMD, which can quite possibly boost the thermopower to such as extent that it more than compensates the loss of mobility in $zT$, opens a completely new approach to optimizing thermoelectrics. Indeed, the large number of known PM semiconductors constitute a hitherto unexplored class of materials that should be investigated to find other high-*zT* semiconductors. Finally, in principle, it should be easier to reduce the lattice thermal conductivity in materials with PMD, because the electronic mobility in those materials is already limited by *s*-*d* scattering to a low number. Alloying or nanostructuring to lower the lattice thermal conductivity will thus have a smaller relative detrimental effect on the electron mobility than it has in non-magnetic systems.

**8. Conclusion**

This article is an attempt to give a didactic introduction to the field of spin caloritronics, which studies mixed spin, heat, and charge transport under the combined effects of a magnetic field gradient, an electrical field, and a temperature gradient. The classical field of thermoelectrics (mixed heat/charge transport) thus is broadened by the number of possible combinations of effects, i.e., a factor 3!/2!=3. The article also reviews a few of basics of magnetism, e.g., magnons, as well as of the tools used to measure and generate spin fluxes, e.g., the SHE and ISHE effects. These tools are as critical to the field of spin transport as voltmeters, current sources, thermometers, and heat sinks are to thermoelectrics. Perhaps this introduction and review will make it possible for the community to develop new ideas and find new materials for solid-state thermal-to-electrical energy conversion. The paper refers to many review articles that, in turn, can guide the reader through the research literature.

**Notes and Acknowledgements**

This article summarizes the notes for two lectures on thermal spin transport given at the international school of physics “Enrico Fermi”, Varenna, in July 2019. The first lecture was on the equilibrium thermodynamics of magnetism and the second on thermally driven spin transport. A third lecture in the series was given by the author on thermal transport in topological materials.

The notes for that are being written as a review article for Reports on Progress in Physics. A fourth lecture was given on measurements errors and error bars in thermoelectric measurements; the notes for that are being written as a review article for Nature Communications.

The author acknowledges support of the OSU Center for Emerging Materials, an NSF MRSEC, under grant number 1420451